\pdfoutput=1

\documentclass[11pt]{article}

\usepackage{graphicx}
\usepackage{color}
\usepackage{epstopdf}

\usepackage{amsmath}
\usepackage{amssymb}
\usepackage{latexsym}

\definecolor{wtmred}{rgb}{0.9,0.0,0}
\definecolor{wtmgreen}{rgb}{0,0.7,0}
\definecolor{wtmblue}{rgb}{0,0,0.9}

\usepackage[pdfstartview={FitH},
pdfpagemode={None},colorlinks=true,pdfhighlight=/O,
linkcolor=wtmblue,citecolor=wtmred,urlcolor=wtmgreen]{hyperref}

\def\sst{\scriptscriptstyle}
\def\ss{\scriptstyle}

\newcommand{\txfc}[2]{{\textstyle{\frac{#1}{#2}}}}

\newcommand{\spwd}[2]{\hspace{#1}\mbox{#2}\hspace{#1}}

\newcommand{\mc}{\mathcal}

\newcommand{\hlf}{\txfc{1}{2}}

\newcommand{\goto}{\rightarrow}

\newcommand{\be}{\begin{equation}}
\newcommand{\ee}{\end{equation}}

\newcommand{\brkt}[2]{\left\langle #1\right|\left.\! #2\right\rangle}

\newcommand{\amp}[2]{
\left\langle\rule{0mm}{#1}\right.\!#2\!\left.\rule{0mm}{#1}\right\rangle}

\newcommand{\bbb}{\begin{eqnarray}}
\newcommand{\eee}{\end{eqnarray}}

\newlength{\slashwd}
\newlength{\txtwd}

\newlength{\ofst}

\begin{document}

\vspace*{3cm}

\begin{center}
{\Large\bfseries Notes on Liouville Theory at \(c\leq 1\)}

\vspace*{1.2cm}

Will McElgin\footnote{\texttt{w-mcelgin@uchicago.edu}} \\

\vspace*{1cm}
Enrico Fermi Inst. and Dept. of Physics \\
University of Chicago \\
5640 S. Ellis Ave., Chicago, IL 60637, USA \\
\vspace*{2cm}
{\large\bfseries Abstract} \\
\vspace*{0.5cm}

\parbox{13.5cm}{The continuation of the Liouville conformal field theory to \(c\leq 1\) is considered. The viability of an interpretation involving a timelike boson which is the conformal factor for two-dimensional asymptotically de Sitter geometries is examined. The conformal bootstrap leads to a three-point function with a unique analytic factor which is the same as that which appears along with the fusion coefficients in the minimal models. A corresponding non-analytic factor produces a well-defined metric on fields only when the central charge is restricted to those of the topological minimal models, and when the conformal dimensions satisfy \(h>(c-1)/24\,\). However, the theories considered here have a continuous spectrum which excludes the degenerate representations appearing in the minimal models. The \(c=1\) theory has been investigated previously using similar techniques, and is identical to a non-rational CFT which arises as a limit of unitary minimal models. When coupled to unitary matter fields, the non-unitary theories with \(c\leq -2\) produce string amplitudes which are similar to those of the minimal string.
}

\vspace*{2cm}

\end{center}

\pagebreak

\section{\large Introduction}

The Liouville conformal field theory (for reviews and early work see~\cite{SeibergLV_90}-\cite{Schomerus_0509}) has been widely investigated for well over two decades, although it was some time before a well-defined three-point function was written down~\cite{Dorn&Otto_9403,Zamolodchikov&Zamolodchikov_9506,Teschner_9507}. Since then there has been very substantial progress in understanding Liouville theory on both closed and open surfaces~\cite{Fateev&Zamolodchikov&Zamolodchikov_0001}-\cite{Ponsot_0309}. However, it has been relatively recently~\cite{Ponsot&Teschner_9911,Teschner_0104,Teschner_0303} that the bulk theory has been shown to constitute a rigorously defined prototype (in the sense of \cite{MooreSeibergCFT_89}) for unitary non-rational conformal field theories. Some of the advances in understanding Liouville theory are due to the fact that it shares many properties with the minimal models~\cite{BPZ_84,Dotsenko&Fateev_85}. These are rational theories with \(c< 1\) and, particularly for the unitary family, are perhaps the most completely understood of all CFTs. The coupling of the minimal models to Liouville theory has been considered~\cite{Seiberg&Shih_0312}-\cite{Basu&Martinec_0509} as an interesting string model which is closely associated with two-dimensional gravity. One somewhat problematic feature of this model, shared by all unitary models coupled to the minimal models, is that the conformal dimensions of the Liouville operators which screen minimal model fields are both discrete and bounded from above. Furthermore, a related theory involving a timelike Liouville field provides a seemingly more tractable spacetime interpretation for string amplitudes, as well as an interesting model of two-dimensional asymptotically de Sitter cosmologies. For these and other reasons, efforts have been made~\cite{Gutperle&Strominger_0301}-\cite{Fredenhagen&Schomerus_0409}
to define a non-rational counterpart to Liouville theory for \(c\leq 1\) through the analytic continuation of correlation functions.

One interesting example of a non-rational CFT that is related to Liouville theory is the \(c=1\) model of \cite{Runkel&Watts_0107} which appears as the \(p\goto\infty\) limit of the unitary \((p,p+1)\) minimal models. It was later shown in \cite{Schomerus_0306} that this model results from the continuation of Liouville theory to \(c=1\). This work also included a discussion of the related theories for \(c<1\). The present work uses some of the techniques of \cite{Schomerus_0306} to treat these theories in more detail. The result has been that non-rational conformal field theories arising from a continuation of Liouville theory appear only at the central charges of the topological minimal models. These are given by
\be\label{topccs}
c=13-6\,(q^{-1}+q)
\ee
where \(q\) is a positive integer. Unlike the \(c=1\) theory of \cite{Runkel&Watts_0107}, for \(q>1\) these theories are necessarily non-unitary. However, they share many features with this model, most significantly that only non-degenerate Virasoro representations appear in the spectrum of fields. Furthermore, in order to define a diagonal (Mobius invariant) metric on fields, these theories also require a specification of the identity operator through the derivative of the continued dimension zero field. This procedure produces a spectrum of fields which appears to be restricted to \(h>(c-1)/24\,\), a range for which a timelike interpretation of the continued Liouville boson is difficult to elucidate. The precise correspondence of these theories to the topological minimal models is an interesting matter that will not be addressed here.

The order of topics covered in these notes is as follows. In section~\ref{lvrev} a short review of Liouville field theory is given, along with some discussion of the zero-mode picture. In section~\ref{tlqm} a discussion of the continuation of the Liouville boson to timelike signature is presented, along with known results about the spectrum of normalizable states in the corresponding quantum mechanics. In section~\ref{bootstrap} some conventions are established and a somewhat detailed review is given of the derivation of the shift relation of the Liouville three-point function through the imposition of crossing symmetry on four-point correlators. In section~\ref{3point} the unique (up to vertex operator rescalings) solution of the shift relations is given for \(c\geq 25\,\), and a corresponding function for \(c\leq 1\) is considered which arises from the continuation of the Liouville shift relations. As discussed in \cite{Zamolodchikov_0505}, this function is closely related to the minimal model three-point function, but does not produce the correct fusion coefficients and turns out not to respect Mobius invariance. In section~\ref{continue} the continuation of the Liouville three-point function to \(c\leq 1\) is derived. The analytic solution to the shift relations for \(c\leq 1\) given in section~\ref{3point} appears along with a non-analytic factor previously introduced in \cite{Schomerus_0306} for \(c=1\,\). This factor is then seen to produce a diagonal two-point function only for the central charges of the topological minimal models (\ref{topccs}), and for primary fields of conformal dimension \(h>(c-1)/24\,\). In section~\ref{strings} a short discussion of string amplitudes involving \(c\geq 25\) and \(c\leq1\) Liouville three-point functions is given. These amplitudes include the vertex operator scalings (``leg factors'') of the minimal string as well as the non-analytic factor associated with the non-rational models considered in these notes. Finally, in section~\ref{interp} it is argued that there does not appear to be a sensible interpretation of the correlators presented here in terms of an interacting timelike boson. Also discussed is the interesting fact that the central charges (\ref{topccs}) which lead to well-defined amplitudes are also those for which the dual potential of the coulomb gas treatment of the \(c\leq 1\) CFT vanishes.

\section{\label{lvrev}\large Liouville Field Theory}

The Liouville conformal field theory on closed surfaces is motivated by the following action:
\be\label{LVact}
S_{\sst L}[\phi](b,\mu)=\frac{1}{4\pi}\int d^{2}\sigma\sqrt{g}
\left((\nabla\phi)^2+QR\phi+4\pi\mu\,e^{2b\phi}\right)
\ee
The central charge of the theory is \(c=1+6Q^2\,\), and \(\mu\) is taken to be real and strictly positive throughout these notes. For the interaction to be a marginal perturbation of the linear dilaton CFT it is required that \(Q=b+b^{-1}\). Note that the action satisfies the relation
\be\label{actionKPZ}
S_{\sst L}[\phi](b,\mu)=S_{\sst L}[\phi+\ln\mu/2b](b,1)
\,-\,Q\chi\ln\mu/2b 
\ee
where \(\chi\) is the Euler number of the surface. As discussed below, this leads to the KPZ scaling relation on correlation functions of primary vertex operators. Defining \(U_{\sst L}=\mu\,e^{2b\phi}\,\), it turns out that crossing symmetry of the Liouville CFT requires the introduction of the following ``dual'' interaction\,:
\be\label{sdpot}
\tilde{U}_{\sst L}=\tilde{\mu}\,e^{2\phi/b}
\ee
where \(\mu\) and \(\tilde{\mu}\) are related by
\be\label{selfdual}
(\pi\mu\gamma(b^2))^{1/b}=(\pi\tilde{\mu}\gamma(b^{-2}))^{b}
\ee
Here \(\gamma(x)=\Gamma(x)/\Gamma(1-x)\,\). That \(U_{\sst L}\) and \(\tilde{U}_{\sst L}\) in~(\ref{sdpot}) are marginal (\(h=\bar{h}=1\)) follows from the fact that for the linear dilaton stress tensor
\be
T=-\partial\phi\partial\phi+Q\,\partial^2\phi
\ee
the conformal dimension of \(e^{2a\phi}\) is \(h_{\sst L}(a)=h_{\sst L}(Q-a)=a(Q-a)\,\). 
The duality symmetry \(\mu\goto\tilde{\mu}\) and \(b\goto b^{-1}\) is an exact symmetry of Liouville correlation functions. Due to the form of \(\tilde{\mu}(\mu,b)\), the interaction \(\tilde{U}_{\sst L}\) also preserves the relation~(\ref{actionKPZ}).

The Liouville CFT on the sphere is characterized by the three-point correlation function
\be\label{3ptfnc}
\left\langle V_{a_3}(z_3)V_{a_2}(z_2)
V_{a_1}(z_1)\right\rangle=\frac{C_{\sst L}(a_3,a_2,a_1)}{
(z_{12}\bar{z}_{12})^{h_1+h_2-h_3}
(z_{23}\bar{z}_{23})^{h_2+h_3-h_1}
(z_{31}\bar{z}_{31})^{h_3+h_1-h_2}}
\ee
Here \(h_j=\bar{h}_j=a_j(Q-a_j)\) is the conformal dimension of the primary vertex operator \(V_{a_j}\). Crossing symmetry and the truncated operator product expansions of the level-two degenerate primary operators \(V_{-b/2}\) and \(V_{-b^{-1}/2}\) lead to difference equations which, for real \(b\), produce a unique solution~\cite{Dorn&Otto_9403,Zamolodchikov&Zamolodchikov_9506,Teschner_9507} for \(C_{\sst L}(a_3,a_2,a_1)\)\,:
\be\label{DOZZ}
C_{\sst L}(a_3,a_2,a_1)=
\left(\pi\mu\gamma(b^2)\,b^{2(1-b^2)}\right)^{(Q-\hat{a})/b}
\frac{\Upsilon_{b}(b)}{\Upsilon_{b}(\hat{a}-Q)}
\prod_j\frac{\Upsilon_{b}(2a_j)}{\Upsilon_{b}(\hat{a}-2a_j)}
\ee
Here \(\hat{a}=\sum_j a_j\) and 
\(\Upsilon_{b}(a)=\Upsilon_{b}(Q-a)=\Upsilon_{b^{-1}}(a)\) is an entire function with zeros at \(a=-nb-m/b\) and \(a=Q+nb+m/b\), with \(n\) and \(m\) non-negative integers. For the strip \(0<a<Q\,\), \(\Upsilon_{b}(a)\) has the following integral representation
\be
\ln\Upsilon_{b}(a)=\int_{0}^{\infty}\frac{dt}{t}
\left[(Q/2-a)^2e^{-2t}-\frac{\sinh^2[(Q/2-a)t]}{\sinh(bt)\sinh(t/b)}\right]
\ee
In fact, \(\Upsilon_{b}(a)\) can be analytically continued to the entire complex \(b^2\) plane except for the negative real axis, a point relevant to the timelike continuation we would like to consider below. Furthermore, \(\Upsilon_{b}(a)\) satisfies
\bbb\label{upsshift}
\Upsilon_{b}(a+b)\!\!& = &\!\!\gamma(ba)\,b^{(1-2ba)}\,\Upsilon_{b}(a) \nonumber\\
\Upsilon_{b}(a+b^{-1})\!\!& = &\!\!\gamma(a/b)\,b^{(2a/b-1)}\,\Upsilon_{b}(a)
\eee
These relations allow the following reflection symmetry to be derived from the three-point function
\be\label{reflect}
V_{a}=R_{\sst L}(a)\,V_{Q-a}
\ee
Here \(R_{\sst L}(a)\) satisfies \(R_{\sst L}(a)R_{\sst L}(Q-a)=1\,\), and is given by
\be\label{refcoef}
R_{\sst L}(a)=
\left(\pi\mu\gamma(b^2)\right)^{(Q-2a)/b}\;
\frac{\gamma(2ab-b^2)}{b^2\,\gamma(2-2a/b+b^{-2})}
\ee
Note that (\ref{reflect}) is written as a relation between vertex operators rather than simply as a symmetry of the three-point function since it also holds for all correlators on surfaces of arbitrary genus, including in the boundary theory~\cite{
Fateev&Zamolodchikov&Zamolodchikov_0001}-\cite{Ponsot_0309}.
Thus there is a single primary vertex operator for a given conformal dimension,\footnote{This assumption is required in order to get a unique answer to the conformal bootstrap.} and the (perturbative) vertex operator with \(h=\bar{h}=a(Q-a)\) may be written in terms of exponentiated free fields as
\be
V_{a}=e^{2a\phi}=R_{\sst L}(a)\,e^{2(Q-a)\phi}
\ee
From the form of the dual cosmological constant \(\tilde{\mu}(\mu,b)\,\), it may be seen that
\be
\frac{\partial\tilde{\mu}}{\partial\mu}=b^{-2}\,\tilde{\mu}/\mu=R_{\sst L}(b)
\ee
and that the form of the dual potential (\ref{sdpot}) is consistent with the reflection property :
\be
\frac{\partial\,}{\partial\mu}\,U_{\sst L}=
\frac{\partial\,}{\partial\mu}\,\tilde{U}_{\sst L}=V_{b}
\ee
The operator equation of motion then reads
\be
-2\nabla^2\phi+QR+8\pi\mu b\,V_{b}=0
\ee
Also implied by the form of \(V_{a}\) is the KPZ scaling relation:
\be\label{kpz}
\left\langle V_{a_n}\ldots V_{a_1}\right\rangle_{\mu}=
\mu^{(Q-\sum_{j}a_j)/b}\;\left\langle V_{a_n}\ldots V_{a_1}\right\rangle_{\mu=1}
\ee
For particular values of the charges, it is possible to compute the correlation functions perturbatively in \(\mu\) and \(\tilde{\mu}\) as follows
\bbb\label{LVpert}\lefteqn{
\left\langle V_{a_n}(z_n)\ldots V_{a_1}(z_1)\right\rangle_{\mu}\,= 
\sum_{q,p=0}^{\infty}\frac{(-1)^{(q+p)}}{q!\,p!}\int d^{2}x_q\ldots d^{2}x_1\int d^{2}y_p\ldots d^{2}y_1 \nonumber}\hspace{3cm}\\ & & 
\left\langle V_{a_n}(z_n)\ldots V_{a_1}(z_1)\,
U_{\sst L}(x_q)\ldots U_{\sst L}(x_1)\,
\tilde{U}_{\sst L}(y_p)\ldots \tilde{U}_{\sst L}(y_1)\right\rangle_{-iQ}
\eee
The correlator \(\left\langle\ldots\right\rangle_{-iQ}\) is that for the linear dilaton CFT and vanishes unless the sum of the charges (coefficients of \(2i\phi\)) in a given product of exponentials equals \(-iQ\). This is a significant restriction on the charges which allow a perturbative computation of the correlators. However, such a calculation precisely reproduces the residues of all of the poles in~(\ref{DOZZ}).

The reflection property~(\ref{reflect}) is related to normalizability of the primary vertex operators in the sense that, as for solutions in the zero-mode quantum mechanics, there is a single vertex operator per conformal dimension. There is a further restriction on the spectrum of normalizable vertex operators, also seen in the zero-mode quantum mechanics, which arises from the two-point function. Defining \(2a_j=Q+ip_j\,\), the two-point function takes the form
\be\label{norm}
\left\langle V_{a_2}(z_2)V_{a_1}(z_1)\right\rangle=
\lim_{a_3\goto 0}\left\langle V_{a_3}(z_3)
V_{a_2}(z_2)V_{a_1}(z_1)\right\rangle=
\frac{2\pi\delta(p_1+p_2)\,+
R_{\sst L}(a_1)\,2\pi\delta(p_1-p_2)}
{(z_{12}\bar{z}_{12})^{2h_1}}
\ee
Use has been made here of the identities (\ref{upsshift}), one consequence of which is \(\Upsilon'_b(0)=\Upsilon_b(b)\,\). The charges of the normalizable vertex operators are thus of the form \(2a_j=Q+ip\), with \({\rm Im}(p)=0\) and \({\rm Re}(p)>0\,\), as allowed by~(\ref{reflect}). These normalizable operators, which satisfy \(h=(Q^2+p^2)/4\geq Q^2/4\,\), comprise the spectrum of states of the spacelike Liouville CFT.  In particular, factorization of correlators of normalizable operators involves only this spectrum as intermediate states. With this choice for the spectrum, the three point function (\ref{DOZZ}) has been shown~\cite{Ponsot&Teschner_9911,Teschner_0104,Teschner_0303} to give rise to a well-defined non-rational unitary CFT.

\subsection*{Liouville Quantum Mechanics}

The above picture can be made more clear by examining the zero-mode quantum mechanics associated with the action~(\ref{LVact}). Taking \(p^2=4h-Q^2\,\), the following zero-mode wave equation applies to solutions \(\psi_{p}\) associated with primary fields of dimension \(h\,\): 
\be\label{SLZM}
\left(-\frac{\partial^2\,}{\partial\phi^2}+
4\pi\mu\,e^{2b\phi}\right)\psi_{p}(\phi)
=p^2\,\psi_{p}(\phi)
\ee
The solutions are the Bessel functions
\be\label{ifnc}
{\rm I}(ip/b,e^{b\phi}\sqrt{4\pi\mu/b^2})\,\simeq\, e^{ip\phi}\;
\frac{\left(\pi\mu/b^2\right)^{ip/2b}}
{\Gamma(1+ip/b)}
\ee
where the behavior in the region \(\phi\goto -\infty\) has been shown. 
Defining \(x=e^{b\phi}\sqrt{4\pi\mu/b^2}\,\), 
as \(\phi\goto \infty\) these solutions have the asymptotic form 
\be\label{iasym}
{\rm I}(ip/b,x)
\,\sim\, \frac{e^{-b\phi/2}}{\sqrt{2\pi}}\,
\left(4\pi\mu/b^2\right)^{-1/4}
\left(e^{x}+i\,e^{- p\pi/b}\,e^{-x}\right)
\ee
Only the linear combination
\be\label{kbes}
{\rm K}(ip/b,x)
=i\,\frac{\pi}{2}\,
\frac{{\rm I}(+ip/b,x)-
{\rm I}(-ip/b,x)}{\sinh(\pi p/b)}
\ee
vanishes at large \(\phi\,\). This choice corresponds to the reflection property~(\ref{reflect}) in the CFT. In particular, taking 
\(2a=Q+ip\),~(\ref{refcoef}) may be written as
\be\label{reflect2}
R_{\sst L}(a)=
-\left(\pi\mu\gamma(b^2)\right)^{-ip/b}\;
\frac{\Gamma(1+ip/b)}{\Gamma(1-ip/b)}\,
\frac{\Gamma(1+ipb)}{\Gamma(1-ipb)}
\ee
which for \(p\) of \(\rm O(1)\) and small \(b\) corresponds with the reflection coefficient associated with the behavior of the \(\rm K\) function as \(\phi\goto -\infty\) (see~(\ref{zmwf}) below). For real \(p\) it may be seen that \(R_{\sst L}(a)\) is a pure phase, as expected for a completely reflecting potential. Furthermore, only for real \(p\) is the \(\rm K\) function normalizable, and thus the zero-mode picture reproduces the spectrum implied by the two-point function~(\ref{norm}).

\begin{figure}[h]
\label{slfig}
\begin{center}
\vspace{0.5cm}
\begin{picture}(140,140)
\put(113,56){\(Q/2\)}
\put(96,56){\(1\)}
\put(77,56){\(b\)}
\put(134,135){\(a\)}
\includegraphics{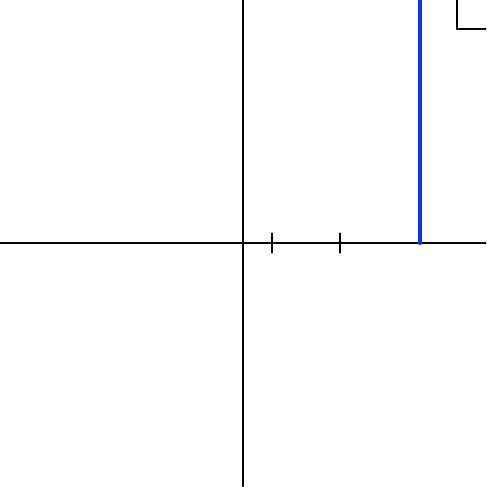}
\end{picture}
\hspace{1.5cm}
\begin{picture}(227,140)
\put(120,140){\(2h\)}
\put(230,10){\(\hat{\phi}\)}
\put(-18,39){\(\txfc{1}{2}Q^2\)}
\put(122,33){\(h=1\)}
\includegraphics{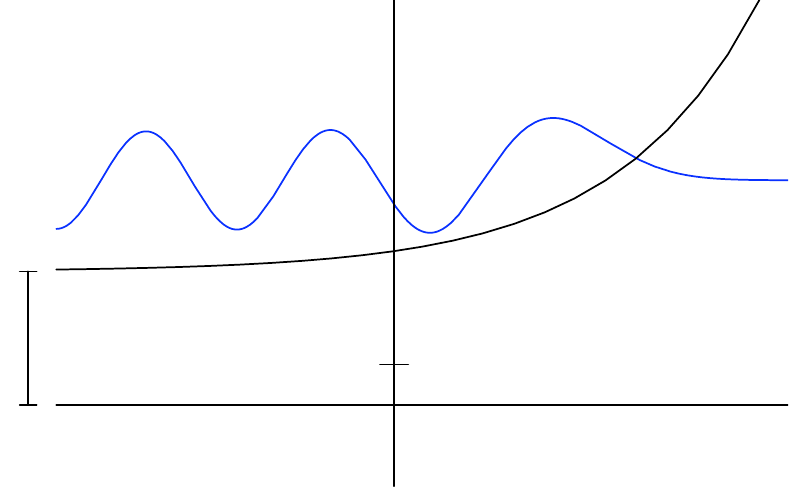}
\end{picture}
\vspace{0.5cm}
\caption{
The figure at left above shows the complex \(a\) plane in spacelike Liouville quantum mechanics, with the spectrum of normalized states shown in \textcolor{blue}{blue}. There is a continuum of states for \(p\in\mathbb{R}_{+}\,\), where \(2a=Q+ip\,\). The figure at right is a plot of the potential \(\exp(2b\hat{\phi})+Q^2/2\,\) for the unit mass Schroedinger equation corresponding to (\ref{SLZM}) with energy \(2h\,\), where the substitution \(2b\hat{\phi}=2b\phi+\ln(2\pi\mu)\,\) has been made. Also shown is a plot of the solution \({\rm K}(ip/b,\exp(b\hat{\phi})\sqrt{2/b^2})\,\) for \(p=10\,b\,\), where \(h=(Q^2+p^2)/4\,\). The choice \(b=0.3\) has been made in both figures.}
\end{center}
\end{figure}
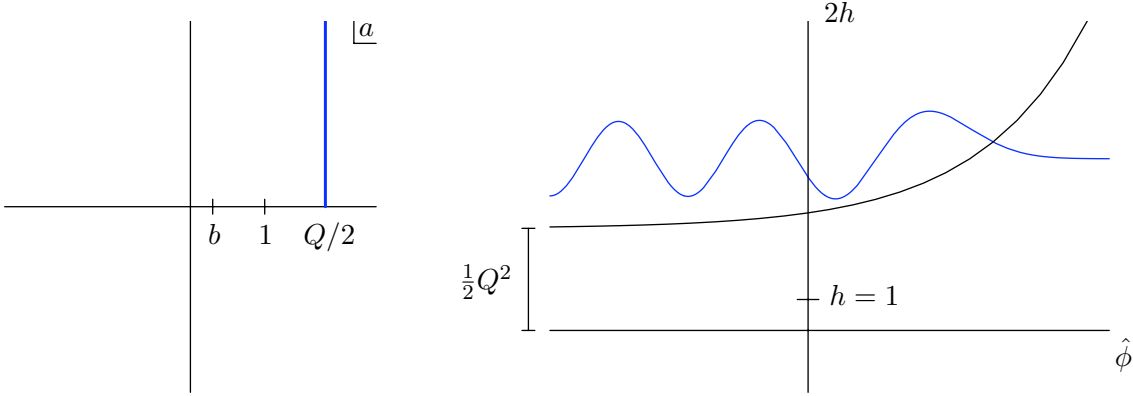

To see that the zero-mode quantum mechanics exhibits the behavior of the three-point function~(\ref{DOZZ}), we define the wavefunctions
\be\label{zmwf}
\Psi(p,\phi)=\frac{2\left(\pi\mu/b^2\right)^{-ip/2b}}
{\Gamma(-ip/b)}\,
{\rm K}(ip/b,x)\,\simeq\,e^{ip\phi}+
\frac{\Gamma(ip/b)}{\Gamma(-ip/b)}\,
\left(\pi\mu/b^2\right)^{-ip/b}\,e^{-ip\phi}
\ee
Where the behavior as \(\phi\goto -\infty\) is shown. 
As \(\phi\goto \infty\,\), \(\Psi(p,\phi)\) has the asymptotic form
\be
\Psi(p,\phi)\,\sim\,
\sqrt{2\pi}\,\frac{\left(\pi\mu/b^2\right)^{-ip/2b}}
{\Gamma(-ip/b)}\,
\left(4\pi\mu/b^2\right)^{-1/4}\,
e^{-b\phi/2}\,\exp\left(-e^{b\phi}\sqrt{4\pi\mu/b^2}\right)
\ee
Defining 
\be\label{msscharges}
2a_1=Q+ib\hat{p}\hspace{1cm} 
2a_2=Q+ib\hat{k}\hspace{1cm} 
2a_3=2b\hat{s} 
\ee
the zero-mode analog of the structure constant is given by
\be\label{mssint}
C_{{\sst L} 0}(a_3,a_2,a_1)=
\int d\phi\,\Psi(b\hat{p},\phi)\,\Psi(b\hat{k},\phi)
\;e^{2b\hat{s}\phi}
\ee
After performing the integral it is found that
\be\label{mss3pt}
C_{{\sst L} 0}(a_3,a_2,a_1)=\frac{(1/b)\,
\left(\pi\mu/b^2\right)^{-(\hat{s}+i(\hat{p}+\hat{k})/2)}}
{\Gamma(-i\hat{p})\,\Gamma(-i\hat{k})\,\Gamma(2\hat{s})}\;
|\Gamma(\hat{s}+i(\hat{p}+\hat{k})/2)|^2\,
|\Gamma(\hat{s}+i(\hat{p}-\hat{k})/2)|^2\,
\ee
Using the asymptotic form of 
\(\Upsilon_b(b\hat{x})\) for small \(b\) 
\be
\Upsilon_b(b\hat{x})\,\sim\,\Upsilon_b(b)\,
b^{(1-\hat{x})}\,(\Gamma(\hat{x}))^{-1}
\ee 
the exact Liouville structure constant~(\ref{DOZZ}) with the charges 
(\ref{msscharges}) coincides with~(\ref{mss3pt}) for small \(b\)
\be
C_{\sst L}(a_3,a_2,a_1)\,\simeq\,C_{{\sst L} 0}(a_3,a_2,a_1)
\ee
Taking the limit \(\hat{s}\goto 0\) for small \(b\), the expression for the two-point function~(\ref{norm}) is recovered from the zero-mode three-point function~(\ref{mss3pt})
\be
\lim_{\hat{s}\goto 0}\,C_{{\sst L} 0}(a_3,a_2,a_1)=
2\pi\delta(p+k)\,+\,
\left(\pi\mu/b^2\right)^{-ip/b}\,
\frac{\Gamma(ip/b)}{\Gamma(-ip/b)}
\,2\pi\delta(p-k)
\ee
where the substitutions \(p=b\hat{p}\) and \(k=b\hat{k}\) have been made.

\section{\label{tlqm}\large Timelike Liouville Quantum Mechanics}

At least naively, a continuation of Liouville theory to timelike signature can be effected by defining \(\phi=i\varphi\) and \(b=-i\beta\). This leads to a timelike analog~\cite{Kobayashi&Tsutsui_9601,Fredenhagen&Schomerus_0308,Hikida_0403} of the Liouville action~(\ref{LVact}) which has the form
\be\label{TLVact}
S_{\sst T}[\varphi](\beta,\rho)=\frac{1}{4\pi}\int d^{2}\sigma\sqrt{g}
\left(-(\nabla\varphi)^2-
\Lambda R\varphi+4\pi\rho\,e^{2\beta\varphi}\right)
\ee
Here \(\rho\) has been substituted for \(\mu\,\), and we have defined \(\Lambda=-iQ=1/\beta-\beta\,\). Choosing the branch \(b=\sqrt{(c-1)/24}-\sqrt{(c-25)/24}\,\), this leads to \(\beta\in(0,1]\subset\mathbb{R}\) for central charge \(c=1-6\Lambda^2\leq1\,\). Taking \(a=-i\alpha\,\), exponential operators \(e^{2a\phi}\) continue to operators 
\(e^{2\alpha\varphi}\) in the timelike theory with conformal dimension 
\(h=\alpha(\Lambda+\alpha)\,\). Besides the change in signature, \(S_{\sst T}\) differs from \(S_{\sst L}\) of~(\ref{LVact}) in at least two other significant ways. First, the term \(QR\phi\) in~(\ref{LVact}) implies that the region of strong string coupling appears as \(\phi\goto\infty\,\), whereas the term \(-\Lambda R\varphi\) in 
(\ref{TLVact}) implies that strong string coupling appears as \(\varphi\goto -\infty\,\). Related to this is the difference between  \(g^{\sst S}_{\mu\nu}=g_{\mu\nu}e^{2a\phi}\) and 
\(g^{\sst T}_{\mu\nu}=g_{\mu\nu}e^{2\beta\varphi}\) when interpreted as metrics of a two-dimensional quantum gravity. The spacelike Liouville equations of motion
\be
-2\nabla^2\phi+QR+8\pi\mu b\,e^{2b\phi}=0
\ee
imply that \(g^{\sst S}_{\mu\nu}\,\), 
for a choice of worldsheet metric \(g_{\mu\nu}\) such that 
\(R[g]=0\), describes a space of constant negative curvature. That is, it describes a locally anti-de Sitter geometry\,:
\be
R[g^{\sst S}]=
e^{-2b\phi}\left(R[g]-2b\nabla^2\phi\right)=-8\pi\mu b^2
\ee
Conversely,~(\ref{TLVact}) implies that \(g^{\sst T}_{\mu\nu}\) describes a space of constant positive curvature; that is, a locally de Sitter geometry~\cite{DaCunha&Martinec_0303,Martinec_0305} with \(R[g^{\sst T}]=8\pi\rho \beta^2\,\). 
Furthermore, the geometric interpretation implies that strong string coupling appears at large scale for the spacelike theory, a regime that the field cannot fully explore due to the form of the potential. For the timelike theory, strong string coupling appears at small spatial scale, a regime that is accessible to the field when coupled to unitary matter which is in a state that is sufficiently excited to overcome the negative Casimir energy of the matter fields.

To attempt to make sense of the CFT correlation functions it is helpful to consider the zero-mode quantum mechanics of~(\ref{TLVact}). After the continuation of~(\ref{SLZM}) using \(\phi=i\varphi\,\), \(b=-i\beta\,\), and \(p=i\omega\,\), the timelike zero-mode Schroedinger equation reads
\be\label{TLZM}
\left(-\frac{\partial^2\,}{\partial\varphi^2}-
4\pi\rho\,e^{2\beta\varphi}\right)\psi_{\omega}(\varphi)
=\omega^2\,\psi_{\omega}(\varphi)
\ee
The sign convention \(p=i\omega\) has been chosen to produce 
\(e^{ip\phi}=e^{-i\omega\varphi}\). Note that the operator \(e^{2\alpha\varphi}\) with 
\(2\alpha=-\Lambda-i\omega\) has conformal dimension \(h=-(\Lambda^2+\omega^2)/4\). 
To find the solutions to~(\ref{TLZM}) it is easiest to recognize that we may continue the solutions to~(\ref{SLZM}). Using \(x=e^{b\phi}\sqrt{4\pi\mu/b^2}\,\) from above, and defining 
\(y=-ix=e^{\beta\varphi}\sqrt{4\pi\rho/\beta^2}\;\), 
the Bessel \({\rm J}\) functions appear as a result of the continuation
\be
{\rm I}(ip/b\,,x)=
{\rm I}(-i\omega/\beta\,,\,iy)=
e^{-\pi\omega/2\beta}\,{\rm J}(-i\omega/\beta\,,\,y)
\ee
The Bessel \({\rm J}\) functions have the 
following \(\varphi\goto -\infty\)  behavior
\be
{\rm J}(i\omega/\beta\,,e^{\beta\varphi}
\sqrt{4\pi\rho/\beta^2}\,)
\,\simeq\, e^{i\omega\varphi}
\;\frac{\left(\pi\rho/\beta^2\right)^{i\omega/2\beta}}
{\Gamma(1+i\omega/\beta)}
\ee
As \(\varphi\goto \infty\) these solutions have the asymptotic form 
\be
{\rm J}(i\omega/\beta,y)
\,\sim\, \frac{e^{-\beta\varphi/2}}{\sqrt{2\pi i}}\,
\left(4\pi\rho/\beta^2\right)^{-1/4}
\left(e^{\omega\pi/2\beta}\,e^{iy}+
i\,e^{-\omega\pi/2\beta}\,e^{-iy}\right)
\ee
For the spacelike theory, the normalizable solutions to the wave equation form a complete orthogonal set. In particular, 
the normalization of \({\rm K}(ip,x)\)~(\ref{kbes}) is given by
\be\label{slint}
\int_0^{\infty}\frac{dx}{x}\;{\rm K}(ip,x)\,{\rm K}(ik,x)=
\frac{\pi}{2p}\,\frac{\pi}{\sinh(\pi p)}\,
\left(\delta(p+k)+\delta(p-k)\right)
\ee
where both \(p\) and \(k\) are real. In the timelike case, the normalizable solutions form an overcomplete set, as may be seen from
\be\label{tlint}
\int_0^{\infty}\frac{dy}{y}\;{\rm J}(s,y)\,{\rm J}(t,y)=
\frac{1}{s+t}\;\frac{\sin(\pi(s-t)/2)}{\pi(s-t)/2}
\ee
which is convergent for \({\rm Re}(s+t)>0\,\). Note that solutions with \((s-t)/2\in\mathbb{Z}_{\neq 0}\) are orthogonal. As expected from the asymptotic form of the solutions, a pole appears at 
\(s=-t\,\). The non-orthogonality of solutions of different energies is associated with the fact that a classical particle will reach \(\varphi\goto\infty\) in finite conformal time. The choice of a self-adjoint Hamiltonian will distinguish a particular orthogonal set of solutions with real energy via boundary conditions at \(\varphi\goto\infty\,\).

Before considering how the requirement of self-adjointness of the Hamiltonian restricts the spectrum of states for timelike Liouville, a brief review of the relevant general operator theory is provided here. Consider an operator \(A\) on a Hilbert space \(\mc{H}\) which is defined on some dense domain \(\mc{D}(A)\subset\mc{H}\,\). Then there exists a unique adjoint \(A^*\) with domain \(\mc{D}(A^*)\supset\mc{D}(A)\) which satisfies \(\brkt{A^{*}g}{h}=\brkt{g}{Ah}\) for all \(h\in\mc{D}(A)\) and \(g\in\mc{D}(A^*)\). An extension \(B\) of \(A\) is an operator with domain \(\mc{D}(B)\supset\mc{D}(A)\) such that \(Bh=Ah\) for all \(h\in\mc{D}(A)\). It may be seen that 
\be
\mc{D}(A^*)\supset\mc{D}(B^*)\supset\mc{D}(B)\supset\mc{D}(A)
\ee 
An operator \(S\) is symmetric if \(\brkt{Sg}{h}=\brkt{g}{Sh}\) for all \(h\in\mc{D}(S)\) and \(g\in\mc{D}(S)\subset\mc{D}(S^*)\). An operator \(H\) is self-adjoint if it is symmetric and \(\mc{D}(H)=\mc{D}(H^*)\). Given a symmetric operator \(S\,\), the conditions for it to possess a self-adjoint extension \(H\) may be explained by starting with the following significant fact about symmetric operators
\be\label{symdom}
\mc{D}(S^*)=\mc{D}(S)+{\rm Ker}(S^*-i)+{\rm Ker}(S^*+i)
\ee
It may be seen that \({\rm Ran}(S\pm i)\) is a closed subspace of \(\mc{H}\) with orthogonal complement \({\rm Ker}(S^*\mp i)\,\). Thus, given the decomposition  
\(\mc{H}={\rm Ran}(S\pm i)\oplus{\rm Ker}(S^*\mp i)\,\), it follows that 
\be
(S^*-i)\psi=(S-i)\phi-2i\eta_{-}=(S^*-i)(\phi+\eta_{-})
\ee
where \(\psi\in\mc{D}(S^*)\,\), \(\phi\in\mc{D}(S)\,\), 
and \(\eta_{-}\in {\rm Ker}(S^*+i)\,\). 
Thus \(\eta_{+}=\psi-\phi-\eta_{-}\in{\rm Ker}(S^*-i)\,\), 
which is equivalent to~(\ref{symdom}). 
It turns out that \(\brkt{S^*g}{h}=\brkt{g}{S^*h}\) 
for all \(h\in\mc{D}(S^*)\) and \(g\in\mc{D}(S^*)\,\) if and only if 
\(\eta_-=U\eta_+\,\), where 
\(U\) is a unitary transformation. This imposes
\be
{\rm dim}\,{\rm Ker}(S^*+i)={\rm dim}\,{\rm Ker}(S^*-i)
\ee
as a sufficient condition for a symmetric operator \(S\) to possesses 
a family of inequivalent self-adjoint extensions. 
Given a particular \(U\), a self-adjoint extension with
\(H=H^*\) and \(\mc{D}(H)=\mc{D}(H^*)\) may be constructed as
\be\label{sadom}
\mc{D}(H)=
\left\{\psi=\phi+\eta_++\eta_-\right.\left|\,\phi\in\mc{D}(S)\,,\,
\eta_+\in{\rm K}_{+}\,,\,\eta_-=U\eta_+\in{\rm K}_{-}\right\}
\ee
where the abbreviation \({\rm K}_{\pm}={\rm Ker}(S^*\mp i)\,\) has been made.

Given the form of the kinetic term in~(\ref{LVact}), a sufficient condition for the spacelike differential operator of~(\ref{SLZM}) to be given by a symmetric operator \((S,\mc{D}(S))\) is 
\be
\mc{D}(S)=\{\,\psi\in\mc{H}\,|\,S\psi\in\mc{H}
\,,\,\lim_{\phi\goto\pm\infty}\psi(\phi)=0
\,,\,\lim_{\phi\goto\pm\infty}\partial_{\phi}\psi(\phi)=0\,\}
\ee
Of course, replacing \(\phi\) with \(\varphi\,\), the timelike theory shares this condition. Note that this implies that there are no boundary conditions on  \(\mc{D}(S^*)=\{\,\psi\in\mc{H}\,|\,S^*\psi\in\mc{H}\,\}\).
An examination of the subspaces 
\({\rm Ker}(S^*\pm i)\) determines the dimensionality of the space of self-adjoint extensions of the naive zero-mode Hamiltonians of the respective Liouville theories. For the spacelike theory, the four solutions of \(S^*=p^2/b^2=\pm i\) are given by
\({\rm I}(\sigma\,ie^{\pm i\pi/4},x)\,\), where \(\sigma=\pm 1\) indexes the two solutions of~(\ref{SLZM}). It may be seen from the \(\phi\goto\pm\infty\) behavior in~(\ref{ifnc}\,,\,\ref{iasym}), that none of these solutions are normalizable. Thus \({\rm dim Ker}(S^*\pm i)=0\,\), and \(S\) has a unique self-adjoint extension \(H\) for the spacelike theory. Note that in this case the spectrum is purely continuous as expressed in~(\ref{slint}). For the timelike theory~(\ref{TLZM}), the four solutions of \(S^*=\omega^2/\beta^2=\pm i\) are \({\rm J}(\sigma\,ie^{\pm i\pi/4},y)\,\),
where again \(\sigma=\pm 1\). Given the integral~(\ref{tlint}), it may be seen that the only normalizable solutions are of the form
\be
\psi_\pm(y)=\left(\frac{\pi}{\sinh(\pi/\sqrt{2})}\right)^{1/2}\,
{\rm J}(\mp ie^{\pm i\pi/4},y)
\ee
where we have taken \(||\psi_\pm||=1\,\). Thus 
\({\rm dim Ker}(S^*\pm i)=1\,\), and the operator \(S\) for the timelike theory possesses a one parameter family of self-adjoint extensions \(\mc{D}_{\nu}(H)\,\). The unitary operator required in~(\ref{sadom}) is given by \(U_{\nu}\psi_{+}=\exp(2\pi i\nu)\psi_{-}\,\).

\vspace{0.5cm}

\begin{figure}[h]
\label{tlfig}
\begin{center}
\vspace{0.5cm}
\begin{picture}(140,140)
\put(12,56){\(-\Lambda/2\)}
\put(96,56){\(1\)}
\put(77,56){\(\beta\)}
\put(134,135){\(\alpha\)}
\includegraphics{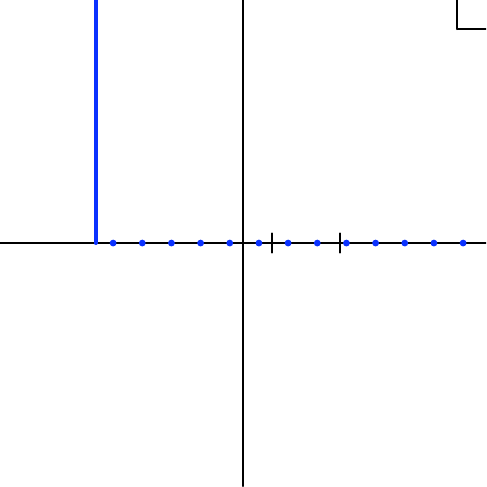}
\end{picture}
\hspace{1.5cm}
\begin{picture}(227,140)
\put(102,140){\(-2h\)}
\put(227,50){\(\hat{\varphi}\)}
\put(-22,76){\(\txfc{1}{2}\Lambda^2\)}
\put(65,44){\(h=1\)}
\includegraphics{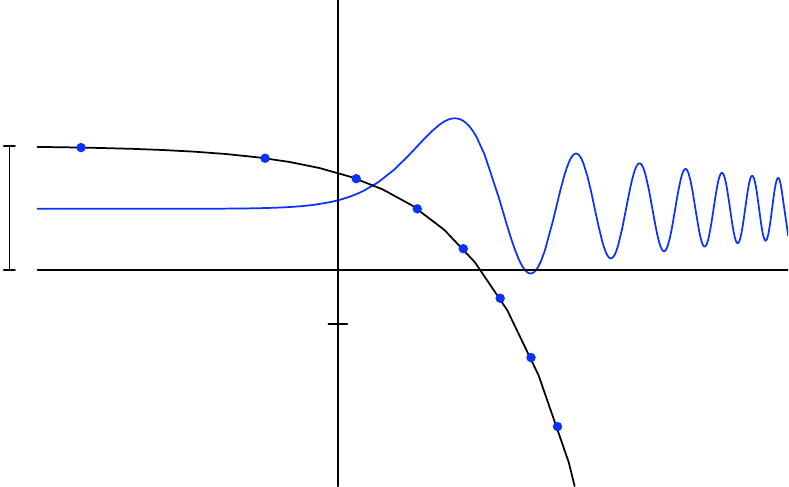}
\end{picture}
\vspace{0.5cm}
\caption{The figure at left above shows the complex \(\alpha\) plane in timelike Liouville quantum mechanics, with the spectrum of normalized states shown in \textcolor{blue}{blue}. For each \(\nu\in(0,1]\,\), which corresponds to a particular self-adjoint extension of the Hamiltonian, there is a continuum of states (\ref{aboveshelf}) with \(\omega\in\mathbb{R}_{+}\,\), where \(2\alpha=-\Lambda+i\omega\,\). In addition, there is a discrete set of states (\ref{belowshelf}) with \(i\omega=2\beta(n+\nu)\) for \(n\in\mathbb{Z}_{\geq 0}\,\). The figure at right is a plot of the potential \(-\exp(2\beta\hat{\varphi})+\Lambda^2/2\,\) for the unit mass Schroedinger equation corresponding to (\ref{TLZM}) with energy \(-2h\,\), where the substitution \(2\beta\hat{\varphi}=2\beta\varphi+\ln(2\pi\rho)\,\) has been made. Also shown is a plot of the discrete solution \({\rm J}(2(n+\nu),\exp(\beta\hat{\varphi})\sqrt{2/\beta^2})\,\) for \(\nu=0.6\,\) and \(n=3\,\), for which \(h=-(\Lambda^2+\omega^2)/4>-\Lambda^2/4\,\). Not shown is an example of the continuum of solutions (\ref{aboveshelf}) corresponding to \(h<-\Lambda^2/4\,\) for this value of \(\nu\,\). The choice \(\beta=0.3\) has been made in both figures.
}
\end{center}
\end{figure}
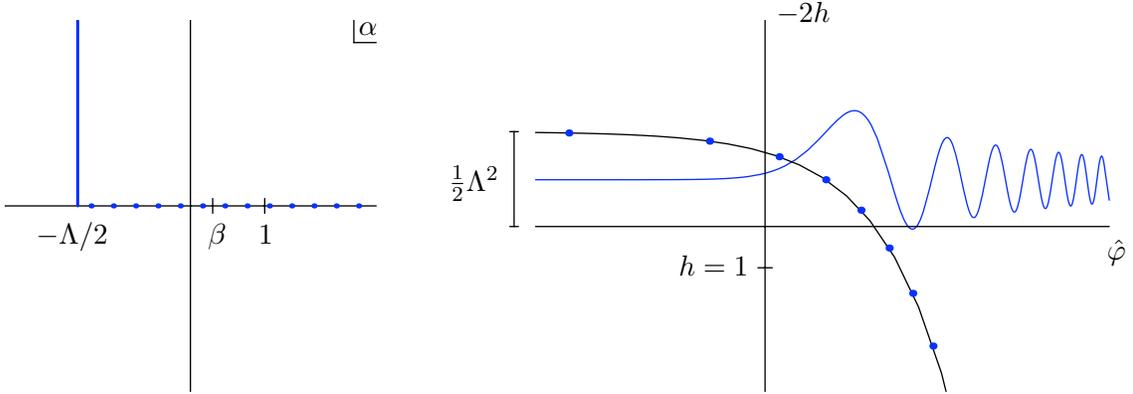

\vspace{0.5cm}

Perhaps the simplest way to determine the spectrum associated with the parameter \(\nu\) is to note from~(\ref{tlint}) that there is a discrete orthonormal set of solutions with imaginary 
\(\omega(\nu,n)=-2i\beta(n+\nu)\) given by
\be\label{belowshelf}
\Psi^{\nu}_{n}(y)=\sqrt{4(n+\nu)}\;
{\rm J}(2(n+\nu),y)
\ee
Here we take \(n\in\mathbb{Z}_{\geq 0}\) and \(\nu\in (0,1]\,\). For real 
\(\omega\,\), a solution corresponding to 
conformal dimension \(h=-(\omega^2+\Lambda^2)/4\) is given by
\be\label{aboveshelf}
\Phi^{\nu}_{\omega}(y)=A^{\nu}_{\omega}\,
{\rm J}(i\omega/\beta,y)+B^{\nu}_{\omega}\,
{\rm J}(-i\omega/\beta,y)
\ee
Imposing that \(\Psi^{\nu}_{n}\) and \(\Phi^{\nu}_{\omega}\) are orthogonal leads to the condition
\be
A^{\nu}_{\omega}\,\sinh(\pi\omega/2\beta+\pi i\nu)=
B^{\nu}_{\omega}\,\sinh(\pi\omega/2\beta-\pi i\nu)
\ee
Choosing 
\be
B^{\nu}_{\omega}=\left(\frac{\omega/\beta}{\sinh(\pi\omega/\beta)}\right)^{1/2}
\ee
the solutions \(\Phi^{\nu}_{\omega}(y)\) are given the continuum normalization
\be
\int_0^{\infty}\frac{dy}{y}\;
\bar{\Phi}^{\nu}_{\omega_1}(y)\,\Phi^{\nu}_{\omega_2}(y)
=\delta(\omega_2-\omega_1)
\ee
where we have implicitly taken \(\omega_j>0\,\). 
It may be shown that this basis is complete, that is
\be
\sum_{n=0}^{\infty}\,
\bar{\Psi}^{\nu}_{n}(y_1)\,
\Psi^{\nu}_{n}(y_2)\,+\,
\int_0^{\infty}d\omega\;
\bar{\Phi}^{\nu}_{\omega}(y_1)\,
\Phi^{\nu}_{\omega}(y_2)=
\delta(\varphi_1-\varphi_2)
\ee
where, as above, \(y=e^{\beta\varphi}\sqrt{4\pi\rho/\beta^2}\,\). Note that, as for the discrete states with \(h>-\Lambda^2/4\,\), the condition of self-adjointness of the Hamiltonian requires that there is a single continuum normalized state for each \(h<-\Lambda^2/4\,\). The zero-mode three-point function associated with the solutions \(\Psi^{\nu}_{n}\) and \(\Phi^{\nu}_{\omega}\) is treated in~\cite{Fredenhagen&Schomerus_0308}. It is clear from the behavior of the solutions that these amplitudes do not follow from the continuation of the Liouville three-point function~(\ref{mss3pt}). In section~(\ref{continue}) the continuation of the Liouville CFT to \(c\leq 1\) will be treated and an interpretation of the correlators in terms of a spacelike boson will be given. As will be discussed below, there does not appear to exist a CFT corresponding to the timelike model treated in this section. 

\section{\label{bootstrap}\large Conformal Bootstrap: Shift Relations}

The existence of degenerate Virasoro representations permits the derivation of analytic shift relations which lead to unique three-point and two-point functions for primary fields in Liouville theory. This procedure, which makes use of the crossing symmetry of the four-point correlator, and the assumption of one primary vertex operator per conformal dimension, is an example of what is often referred to as the conformal bootstrap~\cite{BPZ_84,MooreSeibergCFT_89}. The conformal Ward identities, and the factorization that is utilized in the bootstrap, then allow all correlation functions on the sphere to be computed from the two-point and three-point amplitudes.  Below a \(z\)-frame primary vertex operator will be denoted by \(V_{a}^{(z)}(z,\bar{z})\,\), with a condensed form \(V_{a}^{(z)}\equiv V_{a}^{(z)}(0,0)\) for operators at the origin. The three-point amplitude for primary fields is characterized by the numbers
\be
C_{\sst L}(a_3,a_2,a_1)=
\amp{3mm}{V_{a_3}^{(u)}V_{a_2}^{(w)}V_{a_1}^{(z)}}
\ee
Here \(z\)-frame radial ordering is implicit with the frames related by \(z=1-w=1/u\,\). Symmetry under exchange of the charges \(\{a_j\}\) arises from Mobius symmetry (in this case permutation of the \(z\), \(w\), and \(u\) frames). The two-point function is similarly defined by
\be
G_{\sst L}(a_2,a_1)=\amp{3mm}{V_{a_2}^{(u)}V_{a_1}^{(z)}}
\ee
Before proceeding to the bootstrap, it is necessary to adopt some conventions and make an assumption about the nature of the spectrum. Here the primaries are taken to satisfy a reflection relation \(V_{a}=R_{\sst L}(a)V_{Q-a}\,\), which implies \(R_{\sst L}(a)R_{\sst L}(Q-a)=1\,\). As above, \(h=\bar{h}=a(Q-a)\,\), with \(c=1+6Q^2\,\) and \(Q=1/b+b\,\). For the moment, the branch \(b\in(0,1)\subset\mathbb{R}\) is chosen with \(c>25\), along with values for \(a\) such that \(h\) is real. Given the identification \(2a=Q+ip\,\), elements of the spectrum \(\mathbb{S}\) will be taken to correspond to \(p\in\mathbb{R}_{+}\). The two-point function then takes the following general form 
\be\label{metric}
G_{\sst L}(a_2,a_1)=
R_{\sst L}(a_1)D_{\sst L}(Q-a_1)\,2\pi\delta(p_1+p_2)\,+\,
D_{\sst L}(a_1)\,2\pi\delta(p_1-p_2)
\ee
Here \(D_{\sst L}(a)\) may be seen to satisfy \(R_{\sst L}(a)D_{\sst L}(Q-a)=R_{\sst L}(Q-a)D_{\sst L}(a)\). Mobius invariance implies that the four-point function is characterized by the cross-ratio \(\eta=(z_{12}z_{34})/(z_{13}z_{24})\) as follows 
\bbb
\mc{G}_{a_4 a_3 a_2 a_1}(\eta,\bar{\eta})&=&
\amp{3mm}{V_{a_4}^{(u)}V_{a_3}^{(w)}V_{a_2}^{(z)}(\eta,\bar{\eta})V_{a_1}^{(z)}} \nonumber \\
&=&
\amp{3mm}{V_{a_4}^{(u)}V_{a_1}^{(w)}V_{a_2}^{(w)}(\eta,\bar{\eta})V_{a_3}^{(z)}} \nonumber \\
&=&
\amp{3mm}{V_{a_1}^{(u)}V_{a_3}^{(w)}V_{a_2}^{(u)}(\eta,\bar{\eta})V_{a_4}^{(z)}}
\eee
This leads to the crossing symmetry relation
\be\label{cross1}
\mc{G}_{a_4 a_3 a_2 a_1}(\eta,\bar{\eta})=
\mc{G}_{a_4 a_1 a_2 a_3}(1-\eta,1-\bar{\eta})=
(\eta\bar{\eta})^{-2h_2}\,\mc{G}_{a_1 a_3 a_2 a_4}(1/\eta,1/\bar{\eta})
\ee
The four-point function is expected to factorize as
\be\label{fact}
\mc{G}_{a_4 a_3 a_2 a_1}(\eta,\bar{\eta})=
\int_{\mathbb{S}(\{a_j\})}\hspace{-3mm}da\;
C_{\sst L}(a_4,a_3,a)\,D_{\sst L}(a)^{-1}\,C_{\sst L}(a,a_2,a_1)
\,\left|\mc{F}_{a_4 a_3 a_2 a_1}(a|\eta)\right|^2
\ee
Here the integral extends over a domain \(\mathbb{S}(\{a_j\})\), which depends on the charges and may include discrete contributions. However, if all of the charges are in the spectrum (\(a_j\in\mathbb{S}\)), then \(\mathbb{S}(\{a_j\})=\mathbb{S}\,\). That is, taking \(2a=Q+ip\,\), the integral is over the contour \(p\in\mathbb{R}_{+}\,\). If some charges are outside \(\mathbb{S}\,\), the amplitude may be defined through analytic continuation, with discrete contributions appearing as poles cross the contour of integration. The s-channel conformal blocks \(\mc{F}_{a_4 a_3 a_2 a_1}(a|\eta)\) are holomorphic in \(\eta\) and are determined entirely by conformal invariance. They are related to the t-channel conformal blocks by a fusing matrix \(\sf F\) as follows
\be\label{fuse}
\mc{F}_{a_4 a_3 a_2 a_1}(a|\eta)=
\int_{\mathbb{S}(\{a_j\})}\hspace{-3mm}d\hat{a}\;
{\sf F}{\left[\rule{0mm}{3mm}\right.\hspace{-2mm}
\begin{array}{cc} {\ss a_3}&\hspace{-2mm}{\ss a_2} \\[-2 mm] 
{\ss a_4}&\hspace{-2mm}{\ss a_1} 
\end{array}\hspace{-2mm}
\left.\rule{0mm}{3mm}\right]}(a,\hat{a})\,\mc{F}_{a_4 a_1 a_2 a_3}(\hat{a}|1-\eta)
\ee

It is conventional in Liouville theory to use vertex operator rescalings \(V_{a}\goto f(a)V_{a}\) to set \(D_{\sst L}(a)=R_{\sst L}(a)\,\). This leads to \(R_{\sst L}(a)D_{\sst L}(Q-a)=1\,\), which brings the two-point function~(\ref{metric}) to the form 
\be\label{metric2}
G_{\sst L}(a_2,a_1)=2\pi\delta(p_1+p_2)\,+\,R_{\sst L}(a_1)\,2\pi\delta(p_1-p_2)
\ee
Under rescaling, 
\be
D_{\sst L}(a)\goto(f(a))^2\,D_{\sst L}(a)
\spwd{1cm}{and} R_{\sst L}(a)\goto\frac{f(a)}{f(Q-a)}\, R_{\sst L}(a)
\ee
The choice \(D_{\sst L}(a)=R_{\sst L}(a)\,\) thus restricts further rescalings  to have \(f(a)f(Q-a)=1\,\). In addition, it it convenient to scale the \(a=0\) operator so that it corresponds to the identity (\(V_{0}=\mathbf{1}\)). In this case \(G_{\sst L}(a_2,a_1)=C_{\sst L}(0,a_2,a_1)\,\), and further vertex operator rescalings are restricted to have \(f(0)=1\,\). 

\subsection*{Derivation of the shift relations for \(c>25\)}

In general the integrals in~(\ref{fact}\,,\,\ref{fuse}) are over an infinite number of discretely and continuously indexed Virasoro representations. However, among the non-normalizable operators in Liouville theory are a discrete set of degenerate primary fields for which these expansions are truncated to a finite number of terms. In particular, for the vertex operators \(V_{a(m,n)}\,\), where
\be\label{degvir}
2a(m,n)=-(n-1)\,b\,-(m-1)\,b^{-1}\hspace{1cm}m,n\in\mathbb{Z}_{>0}
\ee
there exists a null descendant at level \(mn\). In the case of \(a(1,2)=-b/2\,\), the assumption of the decoupling of this null descendant results in 
\be\label{null}
\left(\mc{L}_{-2}+b^{-2}\mc{L}^2_{-1}\right)V_{-b/2}=0
\ee
This may be shown to imply that the three-point amplitude \(C_{\sst L}(a_1,-b/2,a_3)\) vanishes unless \(a_3=a_1\pm b/2\), and that the factorization of the four point function~(\ref{fact}) is a sum of two terms. Specifically, the vanishing~(\ref{null}) of the null vector at level 2 in a four-point function with degenerate primary \(V_{a_2}=V_{-b/2}\,\) results in the differential equation
\be\label{diffeq}
\left(-\frac{1}{b^2}\frac{d^2\,}{d\eta^2}+
\left(\frac{1}{\eta}-\frac{1}{1-\eta}\right)\frac{d\,}{d\eta}\,
-\frac{h_1}{\eta^2}-\frac{h_3}{(1-\eta)^2}-
\frac{(h_1+h_2+h_3-h_4)}{\eta(1-\eta)}
\right)\mc{G}_{a_4 a_3 a_2 a_1}(\eta,\bar{\eta})=0
\ee
Here \(h_2=h(-b/2)=-1/2-3b^2/4\,\). The truncated OPE for \(V_{-{\sst\frac{1}{2}}b}\) results in the factorization
\be\label{sfact}
\mc{G}_{a_4 a_3 a_2 a_1}(\eta,\bar{\eta})=
\sum_{a=a_1\pm b/2}\;C_{\sst L}(a_4,a_3,a)\,C_{\sst L}(Q-a,-b/2,a_1)
\,\left|\mc{F}_{a_4 a_3 a_2 a_1}(a|\eta)\right|^2
\ee
Where the normalization \(R_{\sst L}(Q-a)=D_{\sst L}(a)^{-1}\) has been chosen as in~(\ref{metric2}). The differential equation~(\ref{diffeq}) then implies that the conformal blocks in~(\ref{sfact}) are expressed in terms of hypergeometric functions
\be
\mc{F}_{a_4 a_3 a_2 a_1}(a_1\pm b/2\,|\eta)=
\left(\eta\right)^{h(a_1\pm b/2)-h_1-h_2}\,\left(1-\eta\right)^{h(a_3\pm b/2)-h_3-h_2}\,
F(\alpha_\pm,\beta_\pm;\rho_\pm|\eta)
\ee
Where, defining \(2a_j=Q-\lambda_j\,\), we have
\bbb
2\alpha_{\pm}
\!\!&=&\!\! 1\pm b\left(\lambda_1+\lambda_3+\lambda_4\right)\nonumber\\
2\beta_{\pm}
 \!\!&=&\!\! 1\pm b\left(\lambda_1+\lambda_3-\lambda_4\right)\\
\rho_{\pm}
\!\!&=&\!\! 1\pm b\,\lambda_1\nonumber
\eee
Taking \(\,s,t=\pm 1\,\), and defining
\be
{\sf F}_{st}={\sf F}{\left[\rule{0mm}{3mm}\right.\hspace{-2mm}
\begin{array}{cc} {\ss a_3}\hspace{-1mm}&\hspace{-2mm}{\ss -b/2} \\[-2 mm] 
{\ss a_4}\hspace{-1mm}&\hspace{-2mm}{\ss a_1} 
\end{array}\hspace{-2mm}
\left.\rule{0mm}{3mm}\right]}(a_1+sb/2,a_3+tb/2)
\ee
it may be seen that the elements of the fusing matrix~(\ref{fuse}) are given by
\be
{\sf F}_{\pm\pm}=
\frac{\Gamma(\rho_\pm)\Gamma(\rho_\pm-\alpha_\pm-\beta_\pm)}
{\Gamma(\rho_\pm-\alpha_\pm)\Gamma(\rho_\pm-\beta_\pm)}
\ee
and
\be
{\sf F}_{\pm\mp}=
\frac{\Gamma(\rho_\pm)\Gamma(\alpha_\pm+\beta_\pm-\rho_\pm)}
{\Gamma(\alpha_\pm)\Gamma(\beta_\pm)}
\ee
This computation requires the identities
\bbb
F(\alpha,\beta;\rho\,|\eta)&=& 
\frac{\Gamma(\rho)\Gamma(\rho-\alpha-\beta)}{\Gamma(\rho-\alpha)\Gamma(\rho-\beta)}\;
F(\alpha,\beta;\alpha+\beta-\rho+1\,|1-\eta)\\
&+& \frac{\Gamma(\rho)\Gamma(\alpha+\beta-\rho)}{\Gamma(\alpha)\Gamma(\beta)}\;
(1-\eta)^{\rho-\alpha-\beta}\;F(\rho-\alpha,\rho-\beta;\rho-\alpha-\beta+1\,|1-\eta)\nonumber
\eee
and
\be
F(\alpha,\beta;\rho\,|\eta)=(1-\eta)^{\rho-\alpha-\beta}\;F(\rho-\alpha,\rho-\beta;\rho\,|\eta)
\ee
It is helpful to introduce the definitions
\be
\mc{F}_{\pm}=\mc{F}_{a_4 a_1 a_2 a_3}(a_3\pm b/2|1-\eta)
\ee
and
\be\label{cpmdef}
C_{\pm}(a)=C_{\sst L}(Q-(a\pm b/2),-b/2,a)
\ee
where a sometimes useful relation is
\be
C_{+}(a)=\frac{R_{\sst L}(a)}{R_{\sst L}(a+b/2)}\,C_{-}(Q-a)
\ee
Imposing the crossing symmetry relation~(\ref{cross1}), the fusion relation~(\ref{fuse}) leads to
\bbb\label{cross2}\lefteqn{
\sum_{s=\pm}C_{\sst L}(a_4,a_3,a_1+sb/2)\,C_s(a_1)
\left|{\sf F}_{s+}\mc{F}_{+}+{\sf F}_{s-}\mc{F}_{-}\right|^2\,=
\nonumber } \hspace{5cm} \\
& & \sum_{t=\pm}C_{\sst L}(a_4,a_1,a_3+tb/2)\,C_t(a_3)
\left|\mc{F}_{t}\right|^2
\eee

The vanishing of cross terms in the right side of~(\ref{cross2}) yields
\be\label{shift1}
\frac{C_{\sst L}(a_4,a_3,a_1+b/2)\,C_{+}(a_1)}{C_{\sst L}(a_4,a_3,a_1-b/2)\,C_{-}(a_1)}=
-\,\frac{{\sf F}_{-+}\bar{{\sf F}}_{--}}{{\sf F}_{++}\bar{{\sf F}}_{+-}}=
\frac{\gamma(\rho_{-})\,\gamma(\alpha_{+})\,\gamma(\rho_{+}-\alpha_{+})}
{\gamma(\rho_{+})\,\gamma(\beta_{-})\,\gamma(\rho_{-}-\beta_{-})}
\ee
To solve for the shift relations it is probably most straightforward\footnote{For various techniques and normalizations, see \cite{Teschner_9507,Zamolodchikov_0505,Pakman_0601}.} to introduce a set of vertex operators \({\sf \hat{V}}_{a}\) which satisfy \({\sf \hat{R}}_{\sst L}(a)=\pm 1\,\) and \({\sf \hat{D}}_{\sst L}(a)=R_{\sst L}(0)\,\). This choice, which preserves \({\sf \hat{V}}_{0}=V_{0}=\mathbf{1}\,\), is given by
\be\label{LVvorn}
{\sf \hat{V}}_{a}=
\left(\frac{R_{\sst L}(0)}{R_{\sst L}(a)}\right)^{1/2}\,V_{a}
\ee
It should be noted that this rescaling preserves the form of the left hand side of~(\ref{shift1}), with \({\sf \hat{C}}_{\pm}(a)\) defined for \({\sf \hat{V}}_{a}\,\) as in (\ref{cpmdef}). It will be seen below that, for real \(b\in(0,1]\subset\mathbb{R}\,\), the sign in \({\sf \hat{V}}_{a}=\pm{\sf \hat{V}}_{Q-a}\) depends only on the central charge and not on the charge \(a\,\). Choosing \(a_4=-b/2\,\) and \(a_3=a_1=a\,\), in this normalization we find\footnote{Repeated use is made here and below of the identity \(\gamma(x+1)=-x^2\gamma(x)\,\).}
\be\label{shiftsqr}
\left(\frac{{\sf \hat{C}}_{+}(a)}{{\sf \hat{C}}_{-}(a)}\right)^2=\left(\frac{{\sf \hat{C}}_{\sst L}(-b/2,a,a+b/2)}{{\sf \hat{C}}_{\sst L}(-b/2,a,a-b/2)}\right)^2=
\frac{\gamma(2ab-b^2)\,\gamma(2-2ab+2b^2)}
{\gamma(2ab)\,\gamma(2-2ab+b^2)}
\ee
It should be noted that the preceding argument depends only on the conformal dimensions \(h_{j}=a_{j}(Q-a_{j})\) of the operators. Thus\footnote{When it is instructive, and to facilitate comparison of the \(c>25\) and \(c<1\) theories, an argument for the coupling will be added as follows\,: \(R_{\sst L}(a)=R_{\sst L}(a|b)\,\).}
\be\label{cbinv}
{\sf \hat{C}}_{\sst L}(a_3,a_2,a_1|b)=
{\sf \hat{C}}_{\sst L}(a_3,a_2,a_1|b^{-1})
\ee
since the explicit dependence of the three-point function on \(b\) is only through \(Q=b^{-1}+b\,\).  Plugging~(\ref{shiftsqr}) back into~(\ref{shift1}), renaming \(a_4\goto a_2\,\), and taking \(a_1\goto a_1+b/2\,\), leads to the following shift relation for the \({\sf \hat{R}}_{\sst L}(a)=\pm 1\,\) normalization
\be\label{shift2}
\frac{{\sf \hat{C}}_{\sst L}(a_3,a_2,a_1+b)}
{{\sf \hat{C}}_{\sst L}(a_3,a_2,a_1)}=
\left(\frac{\gamma(2a_{1}b)\,\gamma(2a_{1}b+b^2)}
{\gamma(2-2a_{1}b)\,\gamma(2-2a_{1}b+b^2)}\right)^{1/2}\;
\frac{\gamma(\hat{a}b-2a_{1}b-b^2)\,\gamma(2-\hat{a}b+b^2)}
{\gamma(\hat{a}b-2a_{2}b)\,\gamma(\hat{a}b-2a_{3}b)}
\ee
where, as above, \(\hat{a}=\sum_{j}a_{j}\,\). From~(\ref{degvir}), there exists a primary vertex operator with \(a(2,1)=-b^{-1}/2\,\) which also has a null descendant at level 2, thus producing the differential equation~(\ref{diffeq}) with \(b\goto b^{-1}\,\) and \(h_2=h(-b^{-1}/2)\). The associated truncated operator product expansion implies a shift relation for \({\sf \hat{C}}_{\sst L}(a_1,a_2,a_3|b)\) which follows from~(\ref{shift2}) by taking \(b\goto b^{-1}\,\).

\section{\label{3point}\large Shift Relations to Correlators}

It may be seen that the shift relation (\ref{shift2}) has the solution
\be\label{LV3pthat}
{\sf \hat{C}}_{\sst L}(a_3,a_2,a_1)=A_{\sst L}
\left(b^{2(1/b-b)}\right)^{(Q-\hat{a})}
\frac{\Upsilon_{b}(b)}{\Upsilon_{b}(\hat{a}-Q)}
\prod_j g_{\sst L}(a_j)\frac{\Upsilon_{b}(2a_j)}{\Upsilon_{b}(\hat{a}-2a_j)}
\ee
where
\be
g^2_{\sst L}(a)=\frac{\gamma(2-2a/b+b^{-2})\,\gamma(-b^2)}
{\gamma(2ab-b^2)\,\gamma(2+b^{-2})}
\ee
The number \(A_{\sst L}\) corresponds to the overall scale of the three-point function that is left undetermined by the bootstrap. Under reflection we have
\be\label{LVrefpm}
{\sf \hat{R}}_{\sst L}(a)=
{\rm sign}\left(\frac{b^{-2}\gamma(-b^2)}{\gamma(2+b^{-2})}\right)=(-1)^{[b^{-2}]-1}
\ee
where \([b^{-2}]\) is the largest integer less than \(b^{-2}\), and it has been assumed that \(b\in(0,1]\,\). We would now like to use (\ref{LVvorn}) to return to the set of vertex operators \(V_{a}=R_{\sst L}(a)V_{Q-a}\,\), and choose \(R_{\sst L}(a)\,\) to bring the shift relation~(\ref{shift2}) to a form which is analytic in \(\{a_j\}\) and \(b\,\). However, here and below the (identity preserving) vertex operator rescaling \(V_{a}\goto A^{a/Q}_{\sst L}\,V_{a}\) is implemented. Choosing
\be\label{LVref}
R_{\sst L}(a)=A_{\sst L}\;
\frac{b^{-2}\,\gamma(2ab-b^2)}{\gamma(2-2a/b+b^{-2})}
\ee
the three-point function \(C_{\sst L}\) takes the following form
\be\label{LV3pt}
C_{\sst L}(a_3,a_2,a_1)=
A_{\sst L}\left(b^{2(b^{-1}-b)}\right)^{(Q-\hat{a})}
\frac{\Upsilon_{b}(b)}{\Upsilon_{b}(\hat{a}-Q)}
\prod_j\frac{\Upsilon_{b}(2a_j)}{\Upsilon_{b}(\hat{a}-2a_j)}
\ee
The shift relation for the \(V_{a}\) operators is given by
\be\label{LV3ptshift}
\frac{C_{\sst L}(a_3,a_2,a_1+b|b)}
{C_{\sst L}(a_3,a_2,a_1|b)}=H_{\sst L}(a_3,a_2,a_1|b)
\ee
where,
\be\label{Hfunc}
H_{\sst L}(a_3,a_2,a_1|b)=b^{-4}\,\gamma(2a_{1}b)\,\gamma(2a_{1}b+b^2)\
\frac{\gamma(\hat{a}b-2a_{1}b-b^2)\,\gamma(2-\hat{a}b+b^2)}
{\gamma(\hat{a}b-2a_{2}b)\,\gamma(\hat{a}b-2a_{3}b)}
\ee
Using the property of the three-point function (\ref{cbinv}) that \(C_{\sst L}(a_3,a_2,a_1|b)=C_{\sst L}(a_3,a_2,a_1|b^{-1})\,\), the solution to the shift relation~(\ref{LV3ptshift}) and its \(b\goto b^{-1}\) counterpart is unique up to an \(a_{j}\)-independent rescaling. Uniqueness follows from the fact that the ratio of any other solution to that of~(\ref{LV3pt}) must be periodic in both \(b\) and \(b^{-1}\,\), and thus must be independent of \(a_{j}\) for real (in general non-rational) \(b\).
Setting \(A_{\sst L}=(\pi\mu\gamma(b^2))^{Q/b}\,\) and rescaling \(V_{a}\goto A^{-a/Q}_{\sst L}\,V_{a}\) produces the conventional normalization of the three-point function \(C_{\sst L}\)~(\ref{DOZZ}) and reflection coefficient \(R_{\sst L}(a)\)~(\ref{refcoef}) in Liouville theory. Rescaling the result~(\ref{shiftsqr}) produces
\be\label{cpm}
\frac{C_{+}(a)}{C_{-}(a)}=-\frac{\pi\mu}{\gamma(-b^2)}\,
\frac{\gamma(2ab-b^2-1)}{\gamma(2ab)}
\ee
where \(C_{\pm}(a)\) are defined as in~(\ref{cpmdef}). The expression~(\ref{cpm}) may be seen to also result from a perturbative calculation as in~(\ref{LVpert}). We may also define the ``dual'' expression
\be
\tilde{C}_{\pm}(a)=C_{\sst L}(Q-(a\pm b^{-1}/2),-b^{-1}/2,a)
\ee
Taking \(b\goto b^{-1}\) in~(\ref{shiftsqr}) and defining \(\tilde{\mu}\) as in~(\ref{selfdual}) produces 
\be\label{cpmtld}
\frac{\tilde{C}_{+}(a)}{\tilde{C}_{-}(a)}=
-\frac{\pi\tilde{\mu}}{\gamma(-b^{-2})}\,\frac{\gamma(2ab^{-1}-b^{-2}-1)}{\gamma(2ab^{-1})}
\ee
After the introduction of the ``self-dual'' potential in~(\ref{sdpot}), this may also be seen to be the perturbative result. If the expressions (\ref{cpm}) and (\ref{cpmtld}) are computed perturbatively, they may be plugged into (\ref{shift1}) to derive the shift relations. However, crossing symmetry still requires the identification~(\ref{selfdual}).

\subsection*{Solution to the shift relations for \(c\leq 1\)}

While the expression~(\ref{LV3pt}) is, up to the factor \(A_{\sst L}\,\), the unique solution to (\ref{LV3ptshift}) for real \(b\), for complex \(b\) there are solutions related to \(C_{\sst L}\) by a doubly periodic function of \(b\) and \(b^{-1}\). We expect a timelike Liouville theory to be related to the spacelike theory by \(\beta=ib\) and \(\alpha_j=ia_j\). It turns out that the expression~(\ref{LV3pt}) is not analytic for real \(\beta\) and thus a naive continuation of Liouville correlators to timelike signature is not possible. However it is possible to find a unique solution to the shift equations~(\ref{LV3ptshift}) for real \(\beta\). That is, using \(\Lambda=-iQ=\beta^{-1}-\beta\,\), there is a correlator\footnote{The subscript of \(C_{\sst M}\) denotes the generalized minimal model CFT of \cite{Zamolodchikov_0505}. For the minimal model CFT~\cite{BPZ_84}, \(C_{\sst M}\) may computed perturbatively~(\ref{MMpert}) in spacelike signature, and involves only the primary charges of the degenerate representations for rational \(\beta^2\in\mathbb{R}\). To compare the expressions in these notes with the usual minimal model conventions~\cite{DMS}, the charges \(\alpha_j\) should be multiplied by \(-1\,\), and the identifications \(\alpha_{+}=1/\beta\,\), \(\alpha_{-}=-\beta\,\), and \(\alpha_{0}=\Lambda/2\,\) should be made.}
which satisfies 
\be\label{MM3ptshift}
\frac{C_{\sst M}(\alpha_3,\alpha_2,\alpha_1+\beta\,|\beta)}
{C_{\sst M}(\alpha_3,\alpha_2,\alpha_1|\beta)}=
H_{\sst M}(\alpha_3,\alpha_2,\alpha_1|\beta)
\ee
where
\be\label{eq3ptshift}
H_{\sst M}(\alpha_3,\alpha_2,\alpha_1|\beta)\,\equiv\,
H_{\sst L}(-i\alpha_3,-i\alpha_2,-i\alpha_1|-i\beta)
\ee
Given (\ref{MM3ptshift}) and the relation given by \(\beta\goto -\beta^{-1}\,\), \(C_{\sst M}\) is given by
\be\label{MM3pt}
C_{\sst M}(\alpha_3,\alpha_2,\alpha_1)=
A_{\sst M}\left(\beta^{2(\beta^{-1}+\beta)}\right)^{(-\Lambda-\hat{\alpha})}
\frac{\Upsilon_\beta(\beta-\hat{\alpha}-\Lambda)}{\Upsilon_\beta(\beta)}
\prod_j\frac{\Upsilon_\beta(\beta-(\hat{\alpha}-2\alpha_j))}
{\Upsilon_\beta(\beta-2\alpha_j)}
\ee
where \(\hat{\alpha}=\sum_j\alpha_j\,\) and \(A_{\sst M}\) is the scale undetermined by the shift relations. The associated primaries, which will be denoted by \(W_{\alpha}\,\), satisfy \(W_{0}=\mathbf{1}\,\) and
\be
W_{\alpha}=R_{\sst M}(\alpha)\,W_{-\Lambda-\alpha}
\ee
where
\be\label{MMref}
R_{\sst M}(\alpha)=
A_{\sst M}\;
\frac{\beta^{-2}\,\gamma(-2\alpha\beta+\beta^2)}
{\gamma(2-2\alpha/\beta-\beta^{-2})}
\ee
To demonstrate that~(\ref{MM3pt}) leads to~(\ref{MM3ptshift}) and~(\ref{MMref}) requires
\(\Upsilon_\beta(\alpha)=\Upsilon_\beta(\beta^{-1}+\beta-\alpha)\,\). Note in this regard that 
\(h_{\sst M}(\alpha)=h_{\sst M}(-\Lambda-\alpha)=\alpha(\Lambda+\alpha)\,\). 

Considering the three-point function~(\ref{MM3pt}), with \(2\alpha_j=2ia_j=-\Lambda-p_j\,\), the expression analogous to~(\ref{metric2}) is given by
\be\label{MMmet}
G_{\sst M}(\alpha_2,\alpha_1)=
C_{\sst M}(0,\alpha_2,\alpha_1)=
A_{\sst M}\left(\beta^{2(\beta^{-1}-\beta)}\right)^{p_{+}}
\frac{\Upsilon^2_\beta(\beta)F_{\beta}(p_{+})F_{\beta}(p_{-})}
{\Upsilon_\beta(\beta-p_1)\Upsilon_\beta(\beta-p_2)}
\ee
Where \(p_{\pm}=(p_1\pm p_2)/2\,\), and where \(F_{\beta}(k)=F_{\beta}(-k)\) is given by
\be\label{ffunc}
F_{\beta}(k)=\frac{\Upsilon_\beta(\beta-k)}{\Upsilon_\beta(\beta)}\,
\frac{\Upsilon_\beta(\beta+k)}{\Upsilon_\beta(\beta)}
\ee
Note that no delta functions appear in \(G_{\sst M}(\alpha_2,\alpha_1)\,\), and thus it does not in general vanish for primaries of different conformal dimensions. It should be recognized that while (\ref{MM3pt}) satisfies the same analytic shift relation (\ref{MM3ptshift}) as the Liouville expression (\ref{LV3pt}), the conformal bootstrap derivation of these relations based on the factorization of the four-point function (\ref{fact}) relies on a diagonal two-point function (\ref{metric}). Thus there is no reason to expect (\ref{MM3pt}) to correspond to a conformal field theory for general charges \(\{\alpha_j\}\,\). However, examining the diagonal terms 
\be
G_{\sst M}(\alpha,\alpha)=R_{\sst M}(\alpha)
\spwd{1cm}{and}G_{\sst M}(\alpha,-\Lambda-\alpha)=1
\ee
the similarity to~(\ref{metric2}) is obvious. This suggests a sphere partition function normalized as
\be\label{MMpart}
\amp{2.5mm}{\mathbf{1}}=R_{\sst M}(0)=
A_{\sst M}\,\frac{\beta^{-2}\gamma(\beta^2)}
{\gamma(2-\beta^{-2})}
\ee
This expression vanishes for finite \(A_{\sst M}\) for the topological minimal models with \(\beta^{-2}=q\in\mathbb{Z}_{>1}\,\), but in the \(\beta\goto 1\) limit we have \(\amp{2.5mm}{\mathbf{1}}=A_{\sst M}\).

\subsection*{Comments on the minimal model correlators}
As discussed below, the three-point function of the \((p,q)\) minimal models may be computed from (\ref{MM3pt}). The minimal models have rational \(\beta^2=p/q\) with \(q>p>1\,\) and have a field content comprised of degenerate representations with primary charges
\be\label{MMchrg}
2\alpha(m,n)=-(n-1)\,\beta\,+(m-1)\,\beta^{-1}
\hspace{1cm}
\ee
Here \(m\) and \(n\) are restricted to \(1\leq m<p\) and \(1\leq n<q\), with fields identified under the reflection \((m,n)\goto(p-m,q-n)\,\). Choosing \(A_{\sst M}=(\pi\rho\,\gamma(-\beta^2))^{\Lambda/\beta}\,\) leads to an expression for \(C_{\sst M}\) which is very similar to the conventional Liouville expression (\ref{DOZZ})\,, and a perturbative calculation as in (\ref{LVpert}) leads to the minimal model amplitudes. For the \((p,q)\) minimal models the perturbative result may be seen by defining the vertex operators in terms of the spacelike boson \(\phi=i\varphi\,\) as follows
\be
W_{\alpha}=e^{-2i\alpha\phi}=R_{\sst M}(\alpha)\,e^{2i(\Lambda+\alpha)\phi}
\ee
As for the Liouville case, defining \(U_{\sst M}=\rho\,e^{-2i\beta\phi}\,\),  there is a ``dual'' potential
\be
\tilde{U}_{\sst M}=\tilde{\rho}\,e^{2i\phi/\beta}
\ee
where
\be\label{tlselfdual}
(\pi\rho\,\gamma(-\beta^2))^{1/\beta}=
(\pi\tilde{\rho}\,\gamma(-\beta^{-2}))^{-\beta}
\ee
The perturbative result for the minimal models may be derived from
\bbb\label{MMpert}\lefteqn{
\left\langle W_{\alpha_n}(z_n)\ldots W_{\alpha_1}(z_1)\right\rangle_{\rho}\,= 
\sum_{q,p=0}^{\infty}\frac{(-1)^{(q+p)}}{q!\,p!}\int d^{2}x_q\ldots d^{2}x_1\int d^{2}y_p\ldots d^{2}y_1 \nonumber}\hspace{3cm}\\ & & 
\left\langle W_{\alpha_n}(z_n)\ldots W_{\alpha_1}(z_1)\,
U_{\sst M}(x_q)\ldots U_{\sst M}(x_1)\,
\tilde{U}_{\sst M}(y_p)\ldots \tilde{U}_{\sst M}(y_1)\right\rangle_{\Lambda}
\eee
As for (\ref{LVpert}), the correlator \(\left\langle\ldots\right\rangle_{\Lambda}\) is that for the spacelike linear dilaton CFT and vanishes unless the sum of the charges (coefficients of \(2i\phi\)) in a given product of exponentials equals \(\Lambda\). As discussed below, the expression (\ref{MMpert}) must be augmented by the minimal model fusion rules to produce consistent CFT amplitudes.

To relate the expression (\ref{MM3pt}), when evaluated at the charges (\ref{MMchrg}), to the structure constants of the minimal model CFT , it is helpful to define rescaled vertex operators \({\sf \hat{W}}_{\alpha}\) for which  \({\sf \hat{R}}_{\sst M}(\alpha)=\pm 1\,\). As for the Liouville case 
(\ref{LVvorn}) these are defined by
\be\label{MMvorn}
{\sf \hat{W}}_{\alpha}=
\left(\frac{R_{\sst M}(0)}{R_{\sst M}(\alpha)}\right)^{1/2}\,W_{\alpha}
\ee
which preserves the normalization \({\sf \hat{W}}_{0}=W_{0}=\mathbf{1}\,\). The three-point function then takes the form
\be\label{MM3pthat}
{\sf \hat{C}}_{\sst M}(\alpha_3,\alpha_2,\alpha_1)=
A_{\sst M}\left(\beta^{2(1/\beta+\beta)}\right)^{(-\Lambda-\hat{\alpha})}\,
\frac{\Upsilon_\beta(\beta-\hat{\alpha}-\Lambda)}{\Upsilon_\beta(\beta)}
\prod_j g_{\sst M}(\alpha_j)\,\frac{\Upsilon_\beta(\beta-(\hat{\alpha}-2\alpha_j))}
{\Upsilon_\beta(\beta-2\alpha_j)}
\ee
where
\be\label{gfunc}
g_{\sst M}^2(\alpha|\beta)=
g_{\sst L}^2(-i\alpha|-i\beta)=
\frac{\gamma(2-2\alpha/\beta-\beta^{-2})\,\gamma(\beta^2)}
{\gamma(-2\alpha\beta+\beta^2)\,\gamma(2-\beta^{-2})}
\ee
The amplitude (\ref{MM3pthat}) leads to the reflection relation
\be\label{MMrefpm}
{\sf \hat{R}}_{\sst M}(\alpha)=
{\rm sign}\left(\frac{\beta^{-2}\gamma(\beta^2)}{\gamma(2-\beta^{-2})}\right)
=(-1)^{[\beta^{-2}]-1}
\ee
where \([\beta^{-2}]\) is the largest integer less than \(\beta^{-2}\), and it has been assumed that \(\beta\in(0,1]\,\). With the choice 
\(\amp{2.5mm}{\mathbf{1}}=R_{\sst M}(0)=1\,\), (\ref{MM3pthat}) is the 
three-point function of the generalized minimal model (GMM) introduced in \cite{Zamolodchikov_0505}. For the minimal models with \(\beta^2=p/q\) for \(q>p>1\), the rescaling (\ref{MMvorn}) with \(\alpha(m,n)\) as in (\ref{MMchrg}) is non-singular for \(1\leq m<p\) and \(1\leq n<q\,\). However, (\ref{MMvorn}) diverges for \(m=p\,\) or \(n=q\,\), and vanishes for \(m=0\,\) or \(n=0\,\). For the topological minimal models with \(q>p=1\,\), (\ref{MMvorn}) is non-singular for \(m\in\mathbb{Z}_{>0}\) and \(1\leq n<q\,\). Furthermore, unlike the terms in the numerator of the Liouville three-point function (\ref{LV3pt}), which vanish at the locations \(a(m,n)\) (\ref{degvir}) of the degenerate primaries, the denominator in the corresponding GMM three-point function (\ref{MM3pt}) does not have zeros at the charges \(\alpha(m,n)\) of the minimal models.

We would like to explore whether
\be\label{MMope}
{{\sf \hat{C}}^{(m_3,n_3)}}_{\hspace*{10mm}(m_2,n_2) (m_1,n_1)}\,\equiv\,
{\sf \hat{C}}_{\sst M}(\alpha(m_3,n_3),\alpha(m_2,n_2),\alpha(m_1,n_1))/
\amp{2.5mm}{\mathbf{1}}
\ee
produces the operator product expansion of the minimal models.  As might be expected, using (\ref{MMpart}), it may be seen that (\ref{MMope}) is independent of the number \(A_{\sst M}\,\). To relate (\ref{MMope}) to the (\(q>p>1\)) minimal model structure constants, it is sufficient to set \(n_j=1\) and define the more general result through analytic continuation~\cite{Dotsenko&Fateev_85,Runkel&Watts_0107}. In this case \(2\alpha_j=(m_j-1)\beta^{-1}\) satisfies \(h_j=\alpha_j(\Lambda+\alpha_j)\geq 0\,\). From the identities (\ref{upsshift}), for \(m\in\mathbb{Z}_{>0}\,\) it may be shown recursively that
\be\label{upsrec}
\Upsilon_\beta(\beta-(m-1)\beta^{-1})=
\Upsilon_\beta(\beta)\;\beta^{(m-1)(\beta^{-2}m-1)}\,\Omega(m)
\ee
where
\be
\Omega(m)=\prod_{j=1}^{m-1}\,\gamma(j\beta^{-2})
\ee
Here \(\Omega(m)\,\), with \(\Omega(1)\equiv 1\,\), is non-zero and finite for all \(p>m\geq 1\,\). Defining \(m_{\pm}\equiv(m_1\pm m_2)/2\,\), for \(m_{+}\in\mathbb{Z}_{>0}\) and \(m_{-}\geq 0\,\), repeated use of (\ref{upsrec}) leads to
\be\label{opemetev}
{{\sf \hat{C}}^{(1,1)}}_{\hspace*{5mm}(m_2,1) (m_1,1)}=
\frac{\beta^2\gamma(m_{-})}{\gamma(m_{-}\beta^{-2})}\,
\frac{g_{(m_1,1)}g_{(m_2,1)}}{\beta^{4m_{-}}g^2_{(m_{+},1)}}\,
\frac{\Omega^2(m_{+})\,\Omega^2(m_{-}+1)}{\Omega(m_1)\,\Omega(m_2)}
\ee
where \(g_{(m,n)}=g_{\sst M}(\alpha(m,n))\) has been defined. The expression (\ref{opemetev}) equals 1 when \(m_1=m_2\,\), and may be seen to vanish when \(m_{1}\neq m_{2}\,\). It may also be shown that
\be\label{upsrec2}
\Upsilon_\beta(\beta-(m-\hlf)\beta^{-1})=
\Upsilon_\beta(\hlf\beta^{-1})\;
\beta^{(m-1)(\beta^{-2}(m+1)-1)}\,\hat{\Omega}(m)
\ee
where \(m\in\mathbb{Z}_{>0}\,\), and
\be
\hat{\Omega}(m)=\prod_{j=1}^{m-1}\,\gamma((j+\hlf)\beta^{-2})
\ee
Again, the function \(\hat{\Omega}(m)\,\), with \(\hat{\Omega}(1)\equiv 1\,\), is non-zero and finite for all \(p>m\geq 1\,\). For \((2m_{+}-1)/2\in\mathbb{Z}_{>0}\) and \((2m_{-}-1)/2\geq 0\,\), 
\be\label{opemetod}
{{\sf \hat{C}}^{(1,1)}}_{\hspace*{5mm}(m_2,1) (m_1,1)}=
\frac{\beta^{2}}{\beta^{3\beta^{-2}}}\,
\frac{\beta^2\gamma(m_{-})}
{\gamma(m_{-}\beta^{-2})}\,
\frac{g_{(m_1,1)}g_{(m_2,1)}}
{\beta^{4m_{-}}g^2_{(m_{+},1)}}\,
\frac{\hat{\Omega}^2(m_{+}-{\ss\frac{1}{2}})\,
\hat{\Omega}^2(m_{-}+{\ss\frac{1}{2}})}
{\Omega(m_1)\,\Omega(m_2)}\,
\frac{\Upsilon^4_\beta(\hlf\beta^{-1})}
{\Upsilon^4_\beta(\beta)}
\ee
It may be seen that (\ref{opemetod}) does not in general vanish for \(m_{1}\neq m_{2}\,\). Zeros appear in (\ref{MM3pthat}) at the following values of the charges
\be\label{gmmzero}
|\hat{m}-\hat{n}\,p/q|=(2m-1)+(2n-1)\,p/q\hspace{0.5cm}m,n\in\mathbb{Z}_{>0}
\ee
where \(\hat{m}=\sum_{j}m_{j}\,\). Others zeros appear when one of the charges is reflected via \((m_j,n_j)\goto(p-m_j,q-n_j)\,\). This replaces \(|\hat{m}-\hat{n}\,p/q|\) in (\ref{gmmzero}) by \(|\hat{m}_{j}-\hat{n}_{j}\,p/q|\,\), where \(\hat{m}_{j}=\hat{m}-2m_{j}\,\). As mentioned in \cite{Zamolodchikov_0505}, the zeros in the function (\ref{MM3pthat}) form a proper subset of the zeros imposed by the minimal model fusion rules. The latter may be written as
\be
{{\sf F}^{(m_3,n_3)}}_{(m_2,n_2) (m_1,n_1)}=
{{\sf N}^{n_3}}_{n_2n_1}(q)\,{{\sf N}^{m_3}}_{m_2m_1}(p)+
{{\sf N}^{q-n_3}}_{n_2n_1}(q)\,{{\sf N}^{p-m_3}}_{m_2m_1}(p)
\ee
where
\be
{{\sf N}^{n_3}}_{n_2n_1}(q)=\left\{\begin{array}{l}
1\;:\;|n_2-n_1|<n_3<\mbox{min}(n_2+n_1,2q-n_2-n_1)\;,\;
\sum_j n_j\;\;\mbox{odd}\\
0\;:\;\mbox{otherwise}
\end{array}\right.
\ee
The minimal model OPE is then given by
\be\label{mmOPE}
{\sf \hat{W}}^{(w)}_{(m_2,n_2)}
{\sf \hat{W}}^{(z)}_{(m_1,n_1)}\,=\,
\sum_{m_3=1}^{p-1}\sum_{n_3=1}^{q-1}
{{\sf F}^{(m_3,n_3)}}_{(m_2,n_2) (m_1,n_1)}\,
{{\sf \hat{C}}^{(m_3,n_3)}}_{\hspace*{10mm}(m_2,n_2) (m_1,n_1)}
\,\left[\right.{\sf \hat{W}}^{(z)}_{(m_3,n_3)}\left.\right]
\ee
Where, as above, the operators are at the origin of the frames \(z=1-w=1/u\,\), and \(\left[\right.{\sf \hat{W}}_{(m,n)}\left.\right]\) denotes the conformal family of the primary \({\sf \hat{W}}_{(m,n)}\equiv{\sf \hat{W}}_{\alpha(m,n)}\,\). The non-zero result (\ref{opemetod}), and the need to impose the fusion rules by hand in (\ref{mmOPE}), demonstrate that the GMM three-point function is not Mobius invariant, and thus does not give rise to a consistent CFT. It is, however, the unique analytic solution to the shift relation (\ref{MM3ptshift}) and, as in (\ref{mmOPE}), will be present along with a non-analytic coefficient in the \(c\leq 1\) theories considered in the next section.

\section{\label{continue}\large Continuation of Spacelike Amplitudes}

As shown in the last section, the function \(C_{\sst M}\) (\ref{MM3pt}) does not by itself lead to a suitable three-point function unless multiplied by a non-analytic factor which, as in the case of the minimal models, leads to a diagonal (Mobius invariant) two-point function. The question then arises as to whether the continuation of \(C_{\sst L}\) (\ref{LV3pt}) to imaginary \(b\) produces such a factor, and what ranges of central charges and momenta lead to sensible theories. From the shift relations (\ref{LV3ptshift},\,\ref{MM3ptshift},\,\ref{eq3ptshift}), the ratio of \(C_{\sst L}\) to \(C_{\sst M}\) must be related by a doubly periodic function in each of the charges \(a_j\) for \({\rm Im}(b^2)\neq 0\)
\be\label{3ptrat}
\frac{C_{\sst L}(a_3,a_2,a_1|b)}{C_{\sst M}(ia_3,ia_2,ia_1|ib)}
=b^{-1}\,T(a_3,a_2,a_1|b)
\ee
Here \(T(a_j)=T(a_j+b)=T(a_j+b^{-1})\,\) may be computed \cite{Schomerus_0306,Zamolodchikov_0505} from
\be
\Upsilon_b(a)\,\Upsilon_{ib}(ib-ia)=
e^{i\pi(b^{-1}-b+2a)^2/8}\,e^{-\pi i\tau/4}\,
\frac{\vartheta_1(a b^{-1}|\tau)}{\vartheta_3(0|\tau)}
\ee
which is valid for \({\rm Im}(\tau)> 0\), where \(\tau=b^{-2}\,\). Using \(\vartheta_1(0|\tau)=0\) and \(\Upsilon'_b(0)=\Upsilon_b(b)\,\), it may also be shown that
\be
\Upsilon_b(b)\,\Upsilon_{ib}(ib)=b^{-1}e^{i\pi(b^{-1}-b)^2/8}\,
e^{-\pi i\tau/4}\,\frac{\vartheta'_1(0|\tau)}{\vartheta_3(0|\tau)}
\ee
Here the relations~(\ref{upsshift}) and the conventions of \cite{PSKI} for the \(\vartheta\) functions have been used. Since \(T\) is not independent of \(a_j\,\), it follows that \(C_{\sst M}\) is not analytic for real \(b\), where \(C_{\sst L}\) is the unique analytic solution to the shift equations. Similarly, as shown in figure 3, this shows that \(C_{\sst L}\) is not analytic for real \(\beta=ib\). The function \(T\)  in (\ref{3ptrat}) is given by
\be\label{Tfact}
T(a_3,a_2,a_1|b)=e^{-i\pi(Q-2\hat{a})b^{-1}}\,
\frac{\vartheta'_1(0|\tau)}{\vartheta_1((\hat{a}-Q)b^{-1}|\tau)}\prod_{j}
\frac{\vartheta_1(2a_{j}b^{-1}|\tau)}{\vartheta_1((\hat{a}-2a_{j})b^{-1}|\tau)}
\ee
This function may be seen to be analytic in the charges \(a_j\,\) for \({\rm Im}(b^2)< 0\,\). It is also anti-symmetric under reflection in each of its arguments\footnote{This is true for \({\rm Im}(b^2)< 0\,\), but is consistent with the expressions (\ref{LVrefpm}) and (\ref{MMrefpm}), which are valid for \(b^2\in\mathbb{R}\,\).}\,:
\be
\frac{T(Q-a_j)}{T(a_j)}=b^2/\beta^2=-1
\ee

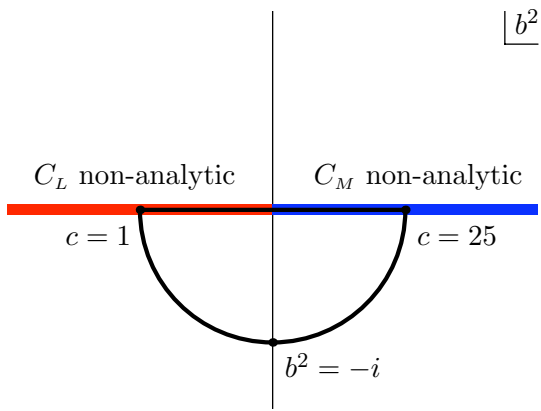
\begin{figure}[h]
\label{bsqrfig}
\begin{center}
\vspace{0.5cm}
\begin{picture}(200,150)
\put(10,85){\(C_{\sst L} \ \mbox{non-analytic}\)}
\put(116,85){\(C_{\sst M} \ \mbox{non-analytic}\)}
\put(105,12){\(b^2=-i\)}
\put(155,62){\(c=25\)}
\put(22,62){\(c=1\)}
\put(192,142){\(b^2\)}
\includegraphics{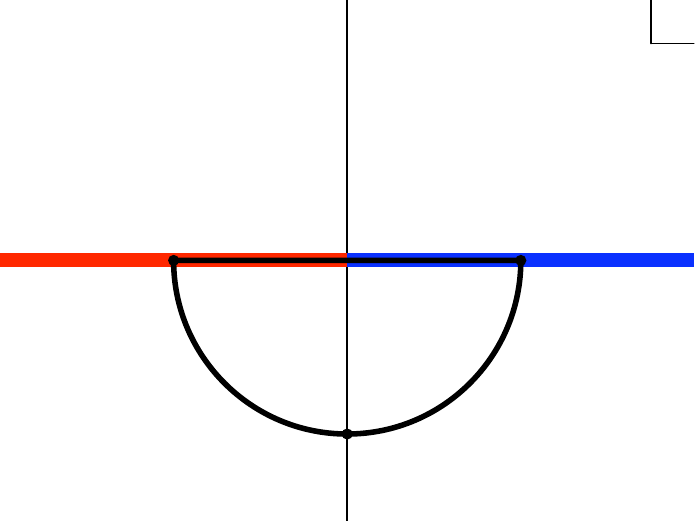}
\end{picture}
\caption{The figure above shows the complex \(b^2\) plane. The black curve is the branch solution of the equation \(c=1+6(b+1/b)^2\) for which \(b\in(0,1]\) for \(c\geq 25\) and \(\beta=ib\in(0,1]\) for \(c\leq 1\). The \textcolor{blue}{blue} half-line (\(b^2\in\mathbb{R}_{+}\)) is the domain on which \(C_{\sst L}\) is the unique analytic solution to (\ref{LV3ptshift}). The \textcolor{red}{red} half-line (\(b^2\in\mathbb{R}_{-}\)) is the domain on which \(C_{\sst M}\) is the unique analytic solution to (\ref{MM3ptshift}). Thus (\ref{eq3ptshift}) requires that \(C_{\sst L}\) and \(C_{\sst M}\) are non-analytic on the \textcolor{red}{red} and \textcolor{blue}{blue} half-lines, respectively.}
\end{center}
\end{figure}

It is helpful in what follows to introduce a number of \(\vartheta_{1}\) function identities. In the conventions used here, the product representation of \(\vartheta_1(x|\tau)\) has the form
\be
\vartheta_1(x|\tau)=2\,e^{\pi i\tau/4}\,\sin(\pi x)\prod_{n=1}^{\infty}
(1-e^{2\pi i n\tau})(1-e^{2\pi i (n\tau+x)})(1-e^{2\pi i (n\tau-x)})
\ee
which leads to
\be
\vartheta'_1(0|\tau)=2\pi\,\eta^3(\tau)=
2\pi\,e^{\pi i\tau/4}\,\prod_{n=1}^{\infty}
(1-e^{2\pi i n\tau})^3
\ee
The \(\vartheta_{1}\) function satisfies the quasi-periodicities
\bbb
\vartheta_1(x+1|\tau) \!&=&\! -\,\vartheta_1(x|\tau)\;=\;
\vartheta_1(-x|\tau) \label{perone}\\[0.1cm]
\vartheta_1(x+\tau|\tau) \!&=&\! 
-\,e^{-2\pi i(x+\tau/2)}\,\vartheta_1(x|\tau) \label{pertau}
\eee
and possesses the following behavior under modular transformations 
\bbb
\vartheta_1(x|\tau+1) \!&=&\! e^{\pi i/4}\,\vartheta_1(x|\tau) \label{tauphase}\\[0.1cm]
\vartheta_1(x/\tau|-1/\tau) \!&=&\! -(-i\tau)^{1/2}\,
e^{\pi ix^2/\tau}\,\vartheta_1(x|\tau)
\eee
Defining \(\hat{x}=\sum_{j=1}^{3}x_{j}\,\), it may be shown that
\be
\vartheta_{1}(2x)\prod_{j}\vartheta_{1}(2x_j)=
\vartheta_{1}(\hat{x}-x)\prod_{j}\vartheta_{1}(\hat{x}-2x_j+x)-
\vartheta_{1}(\hat{x}+x)\prod_{j}\vartheta_{1}(\hat{x}-2x_j-x)
\ee
Here the identification \(\,\vartheta_{1}(x)\equiv\vartheta_{1}(x|\tau)\,\) has been made. A consequence of this identity is 
\be\label{thetaid}
\vartheta'_{1}(0)\prod_{j}\vartheta_{1}(2x_j)=
-\left(\rule{0mm}{5mm}\right.\!
\frac{\vartheta'_{1}(\hat{x})}{\vartheta_{1}(\hat{x})}
-\sum_k \frac{\vartheta'_{1}(\hat{x}-2x_k)}{\vartheta_{1}(\hat{x}-2x_k)}
\!\left.\rule{0mm}{5mm}\right)
\vartheta_{1}(\hat{x})\prod_{j}\vartheta_{1}(\hat{x}-2x_j)
\ee

We would now like to examine the factor \(T\) (\ref{Tfact}) in the limit \({\rm Im}(b^{-2})\goto 0^{+}\). While \(\vartheta_1(x|\tau)\) is extremely singular in this limit, it will be shown below that for \(\beta^2=p/q\) with \((p,q)\) coprime integers which satisfy \(q\geq p\geq 1\,\), the non-analytic factor \(T\) produces non-trivial amplitudes for \(C_{\sst L}\,\). The corresponding central charges
\be\label{tmmcg}
c=13-6\,(p/q+q/p)
\ee
include those of the minimal models. It will further be shown that Mobius invariant amplitudes are produced only for \(c=13-6\,(q^{-1}+q)\,\) of the topological minimal models. These latter models involve degenerate representations with primary charges as in (\ref{MMchrg})\,:
\be
2\alpha(m,n)=-(n-1)/\sqrt{q}\,+(m-1)\,\sqrt{q}
\hspace{1cm}
\ee
but with the restriction \(m\in\mathbb{Z}_{>0}\) and \(1\leq n<q\). While the central charges (\ref{tmmcg}) are rational, the factor \(T\) may be seen to vanish for the degenerate charges \(\alpha(m,n)\)\,, and the amplitude given by \(C_{\sst L}\) for imaginary \(b\) through (\ref{3ptrat}) has a spectrum which does not include these conformal dimensions. The treatment here follows that of \cite{Schomerus_0306} closely, where the \(c=1\) case was considered. For conformal dimensions \(h>(c-1)/24=0\,\), it was found in \cite{Schomerus_0306} that (\ref{3ptrat}) is equivalent to the three-point function of the \(p\goto\infty\) limit of unitary (\(p,p+1\)) minimal models considered in \cite{Runkel&Watts_0107}. This theory involves a continuous set of primary fields which excludes the \(c=1\) degenerate primaries with \(h=n^2/4\) for \(n\in\mathbb{Z}\,\). Taking \(x_j=a_jb^{-1}\) and making use of (\ref{thetaid}) we have
\be
T = -\frac{\vartheta'_{1}(0)}
{\vartheta_{1}(\hat{x})}\prod_{j}
\frac{\vartheta_{1}(2x_j)}
{\vartheta_{1}(\hat{x}-2x_j)} = 
\frac{\vartheta'_{1}(\hat{x})}{\vartheta_{1}(\hat{x})}
-\sum_j \frac{\vartheta'_{1}(\hat{x}-2x_j)}
{\vartheta_{1}(\hat{x}-2x_j)}
\ee
This may be written as
\be\label{Tfact2}
T= \left.\frac{d\,}{dx}\,\ln
\left(\rule{0mm}{5mm}\right.\!
\frac{\vartheta_{1}(x)}
{\prod_j\vartheta_{1}(x-2x_j)}
\!\left.\rule{0mm}{5mm}\right)\right|_{x=\hat{x}}
\ee
As mentioned above, while \(\vartheta_{1}\) exhibits the quasi-periodicities of (\ref{perone}) and (\ref{pertau}), the factor \(T\) in (\ref{Tfact2}) is periodic under \(x_j\goto x_j+1\) and \(x_j\goto x_j+\tau\). This may be seen from
\be\label{thtprd}
\frac{\vartheta'_{1}(x+1)}{\vartheta_{1}(x+1)}=
\frac{\vartheta'_{1}(x)}{\vartheta_{1}(x)}\spwd{1cm}{and}
\frac{\vartheta'_{1}(x+\tau)}{\vartheta_{1}(x+\tau)}=
\frac{\vartheta'_{1}(x)}{\vartheta_{1}(x)}-2\pi i
\ee
or, equivalently, from the fact that \(C_{\sst M}\) and \(C_{\sst L}\,\) satisfy identical shift relations (\ref{LV3ptshift},\,\ref{MM3ptshift},\,\ref{eq3ptshift}) under \(a_j\goto a_j+b\) and \(a_j\goto a_j+b^{-1}\). This implies that the limit \(\,{\rm Im}(b^{-2})\equiv\epsilon\goto 0^{+}\,\) can only lead to a non-trivial three-point function for the choice \(\,\tau=b^{-2}=-r+i\epsilon\,\) with \(r\) a rational number. Otherwise the real periodicities of \(T\) would not have rational ratio in the limit in which \(\vartheta_{1}\) degenerates, and (\ref{Tfact2}) would have to be a constant. This creates a problem since \(C_{\sst M}\) does not by itself lead to a diagonal two-point function. This fact is consistent with \(C_{\sst M}\) being the unique analytic solution to the shift equations for \(\beta=ib\in\mathbb{R}\,\) and leads to the expectation that \(C_{\sst L}\) is a wildly discontinuous function of the central charge for \(c\leq 1\,\).

\subsection*{Correlators for \(h>(c-1)/24\)}

Since for real conformal dimensions the associated momenta must be either real or imaginary, it is useful to express the charges as \(2a_j=Q+ik_j\,\). This leads to
\be\label{pdef}
2x_j=2a_jb^{-1}=2\alpha_j\beta^{-1}=1-q/p-k_j\,\sqrt{q/p}
\ee
where the limit \(\beta=ib\goto \sqrt{p/q}\,\) has been taken. As discussed below, it does not appear that a sensible two-point function exists for the case of imaginary \(k_j\), and thus the three-point function for \(h<(c-1)/24\) will not be treated in these notes. As discussed in \cite{Schomerus_0306}, For \(h_j>(c-1)/24\,\) (\(p_j\in\mathbb{R}\)), the choice \((p=1,q=1)\) leads to the following periodic sawtooth function\,:
\be
\mc{D}_{1}(x)=\lim_{\epsilon\goto 0}\;\frac{\epsilon}{2\pi}\,
\frac{\vartheta'_{1}(x|\tau)}
{\vartheta_{1}(x|\tau)}=\,1/2-(x-[x])
\ee
Here \([x]\) is the largest integer less than \(x\in\mathbb{R}\). From (\ref{thtprd}) it may be seen that this is also the result for the (\(\tau=-q+i\epsilon\)) topological minimal models. For the general \((p,q)\) case this is modified to
\be
\mc{D}_{p}(x)=\lim_{\epsilon\goto 0}\;\frac{\epsilon}{2\pi}\,
\frac{\vartheta'_{1}(x|-q/p+i\epsilon)}
{\vartheta_{1}(x|-q/p+i\epsilon)}=\,\left(1/2-(p\,x-[p\,x])\right)/p
\ee
Note that this expression is independent of \(q\). The general result for \(\mc{D}_{p}\) may be seen to follow from the various \(\vartheta_{1}\) function identities given above, and from the relation
\be
\frac{\vartheta'_{1}(x|\tau)}
{\vartheta_{1}(x|\tau)}=\pi\cot(\pi x)+4\pi
\sum_{n=1}^{\infty}\,\frac{\sin(2\pi n x)}{\exp(-2\pi in\tau)-1}
\ee
It is evident that \(\mc{D}_{p}(x+1/p)=\mc{D}_{p}(x)\,\), and
\be
\frac{\partial\mc{D}_{p}}{\partial x}(x)=
-1+\sum_{n\in\mathbb{Z}}\,\delta(p\,x-n)
\ee
Thus for \(x_j\in\mathbb{R}\) we may write
\be
\lim_{\epsilon\goto 0}\,\frac{\epsilon}{2\pi}\;T\,=\,
\mc{D}_{p}(\hat{x})-\mbox{\(\sum_j\)}\,\mc{D}_{p}(\hat{x}-2x_j)\,=\,
p^{-1}\left(-1+[p\,\hat{x}]-\mbox{\(\sum_j\)}\,[p\,(\hat{x}-2x_j)]\right)
\ee
Using \([-x]=-[x]-1\,\), and defining \(\hat{k}=\sum_{j=1}^{3}k_j\,\), we find the following expression for non-analytic factor in the three-point amplitude for real \(\,k_j\) :
\be\label{Tfact3}
\lim_{\epsilon\goto 0}\,\frac{\epsilon}{2\pi}\;T\,=\,
p^{-1}\left(1+p-q-[\hlf(\hat{k}\,\sqrt{qp}-p+q\,)]+
\mbox{\(\sum_j\)}\,[\hlf((\hat{k}-2k_j)\sqrt{qp}-p+q\,)]\right)
\ee
This may be seen to reduce in the case \((p=1,q=1)\) to the non-analytic coefficient of the three-point function given in \cite{Schomerus_0306}\,, which reproduced the result of \cite{Runkel&Watts_0107} in which the \(c\goto 1\) limit of unitary minimal models was considered. To define the three-point function of the non-rational theory considered here, the limit is taken such that \(A\equiv A_{\sst L}/\epsilon\) is finite. This leads to the following three-point function
\bbb\label{TL3pt}
C_{(p,q)}(\alpha_3,\alpha_2,\alpha_1) &=& 
\lim_{\epsilon\goto 0}\,
C_{\sst L}(-i\alpha_3,-i\alpha_2,-i\alpha_1|-i\beta) \\
&=& 2\pi i\sqrt{q/p}\,A\,
A^{-1}_{\sst M}\,C_{\sst M}(\alpha_3,\alpha_2,\alpha_1|\beta)\;
\lim_{\epsilon\goto 0}\,\frac{\epsilon}{2\pi}\;T
\eee
This results in
\bbb\label{k3pt}
C_{(p,q)}(\alpha_3,\alpha_2,\alpha_1)&=&2\pi i\,p^{-1}\sqrt{q/p}\,A\;
\left(\!\rule{0mm}{4mm}\right.\!-1+[\sqrt{qp}\,\hat{\alpha}]-\mbox{\(\sum_j\)}\,
[\sqrt{qp}\,(\hat{\alpha}-2\alpha_j)]\left.\!\rule{0mm}{4mm}\right)\nonumber \\ & &
\times \left((q/p)^{(q/p-p/q)}\right)^{(1+\hat{\alpha}/\Lambda)}
\frac{\Upsilon_\beta(\beta-\hat{\alpha}-\Lambda)}{\Upsilon_\beta(\beta)}
\prod_j\frac{\Upsilon_\beta(\beta-(\hat{\alpha}-2\alpha_j))}
{\Upsilon_\beta(\beta-2\alpha_j)}
\eee
Here, as above, \(\Lambda=\beta^{-1}-\beta\,\), \(\beta=ib\,\), and \(\alpha_j=ia_j\,\), with \(b^{-2}=-q/p+i\epsilon\,\). 

The two-point function is expected to appear in the limit \(\alpha_3\goto 0\,\). In this case 
\be\label{Tmet}
\lim_{\epsilon\goto 0}\,\frac{\epsilon}{2\pi}\;T
\,=\,p^{-1}\left(\,-1-[\sqrt{qp}\,(\alpha_1-\alpha_2)]-
[\sqrt{qp}\,(\alpha_2-\alpha_1)]\,\right)=0
\ee
This result, which follows since the identity is the primary field of a degenerate representation, clearly does not give rise to a suitable two-point function. The solution given in \cite{Runkel&Watts_0107} is to define 
\be
\mathbf{1}=\lim_{\alpha\goto 0}\frac{\partial\,}{\partial\alpha}\;V_{-i\alpha}
\ee
Thus we define the metric on fields as 
\bbb
G_{(p,q)}(\alpha_2,\alpha_1)&=&
\lim_{\epsilon\goto 0}\,
\lim_{\alpha\goto 0}\,\frac{\partial\,}{\partial\alpha}\,
C_{\sst L}(-i\alpha,-i\alpha_2,-i\alpha_1|-i\beta)\nonumber \\ &=&
2\pi i\sqrt{q/p}\,A\;A^{-1}_{\sst M}\,G_{\sst M}(\alpha_2,\alpha_1|\beta)\;
\lim_{\epsilon\goto 0}\,\frac{\epsilon}{2\pi}\;\lim_{\alpha\goto 0}
\frac{\partial\,}{\partial\alpha}\;T
\eee
Computing the derivative of \(T\,\), we find
\be\label{delta}
\lim_{\epsilon\goto 0}\,\frac{\epsilon}{2\pi}\,
\lim_{\alpha\goto 0}\frac{\partial\,}{\partial\alpha}\,
T(-i\alpha,-i\alpha_2,-i\alpha_1)=
2\sqrt{q/p}\,\sum_{n\in\mathbb{Z}}\,
\left(\rule{0mm}{4mm}\right.\!
\delta(k_{+}\sqrt{qp}-n)-\delta(k_{-}\sqrt{qp}-n)
\!\left.\rule{0mm}{4mm}\right)
\ee
where \(k_{\pm}=(k_1\pm k_2)/2\,\) for \(\,k_j\in\mathbb{R}\,\) as given in (\ref{pdef}). The periodic delta functions appearing in (\ref{delta}) would seem to produce a metric on fields that is not Mobius invariant, since fields of different conformal dimensions could have a non-zero inner product. However, from (\ref{MMmet}),
\be
A^{-1}_{\sst M}\,G_{\sst M}(\alpha_2,\alpha_1|\beta)=
\left((q/p)^{(q/p-p/q)}\right)^{-k_{+}/\Lambda}
\frac{\Upsilon^2_\beta(\beta)F_{\beta}(k_{+})F_{\beta}(k_{-})}
{\Upsilon_\beta(\beta-k_1)\Upsilon_\beta(\beta-k_2)}
\ee
where \(F_{\beta}(k)\) is given by (\ref{ffunc}). This leads to
\be
F_{\beta}(k_{\pm})\,\delta(k_{\pm}\sqrt{qp}-n)=
\delta(k_{\pm}\sqrt{qp}-n)\ 
\frac{\Upsilon_\beta(\beta-n\beta/p)}{\Upsilon_\beta(\beta)}\;
\frac{\Upsilon_\beta(\beta+n\beta/p)}{\Upsilon_\beta(\beta)}
\ee
Thus for the topological minimal models (\(p=1\)), the zeros of the \(\Upsilon\) functions impose
\be
F_{\beta}(k_{\pm})\sum_{n\in\mathbb{Z}}\,
\delta(k_{\pm}\sqrt{q}-n)=\delta(k_{\pm}\sqrt{q})
\ee
and the corresponding metric on fields is in fact diagonal
\be\label{kmet}
G_{q}(\alpha_2,\alpha_1)\equiv G_{(1,q)}(\alpha_2,\alpha_1)=2i\sqrt{q}\,A\,
\left(\rule{0mm}{4mm}\right.\!
2\pi\delta(k_{+})+R_{q}(\alpha_1)\,2\pi\delta(k_{-})
\!\left.\rule{0mm}{4mm}\right)
\ee
Here the reflection coefficient \(R_{q}(\alpha)\) is given by\footnote{Note the minus sign with respect to \(R_{\sst M}\,\)(\ref{MMref})\,, in agreement with the continuation of \(R_{\sst L}\,\)(\ref{LVref}). }
\be\label{refk}
R_{q}(\alpha)=-\frac{\gamma(-2\alpha\beta+\beta^2)}
{\beta^2\,\gamma(2-2\alpha/\beta-\beta^{-2})}=-q\,
\frac{\gamma(-2\alpha/\sqrt{q}+1/q)}
{\gamma(2-2\alpha\sqrt{q}-q)}
\ee
However, for \(p>1\) the two-point function is not diagonal, and thus operators of different conformal dimensions have non-zero inner product. We will take this to mean that only the non-rational theories with central charges 
\be
c=13-6\,(q^{-1}+q)
\ee
lead to well-defined conformal field theories. In the last section some arguments suggesting why this might make physical sense are put forward.\footnote{The significance of the factors of \(i\) in \(C_{q}\equiv C_{(1,q)}\) and \(G_{q}\,\), as well as that of the scaling \(A=A_{\sst L}/\epsilon\,\), is not clear to the author. For the latter at least, the inverse of \(G_{q}\) might be used to raise indices on \(C_{q}\) to obtain finite operator product coefficients in the absence of the scaling.}

\subsection*{Status of the two-point function for \(h<(c-1)/24\)}
Consider the expression 
\be
T=\frac{\vartheta'_{1}(\hat{x}|\tau)}{\vartheta_{1}(\hat{x}|\tau)}
-\sum_j \frac{\vartheta'_{1}(\hat{x}-2x_j|\tau)}
{\vartheta_{1}(\hat{x}-2x_j|\tau)}
\ee
with
\bbb
& & 2x_1=2a_1b^{-1}=1+\tau-i\omega_1\sqrt{q} \\
& & 2x_2=2a_2b^{-1}=1+\tau-i\omega_2\sqrt{q} \\
& & 2x_3=2a_3b^{-1}=2\sigma\sqrt{q}
\eee
Here \(\tau=-q+i\epsilon\), and \(\omega_j\) and \(\sigma\) will be taken to be real. Ultimately we will be interested in the \(\sigma\goto 0\) limit. Defining \(2\omega_{\pm}=\omega_1\pm\omega_2\), and using the \(\vartheta_{1}\) function relations given above we find \(T=\hat{T}(\omega_{+})-\hat{T}(\omega_{-})\), where
\be
\hat{T}(\omega)=\frac{\vartheta'_{1}(\sqrt{q}(\sigma+i\omega)|i\epsilon)}{\vartheta_{1}(\sqrt{q}(\sigma+i\omega)|i\epsilon)}+\frac{\vartheta'_{1}(\sqrt{q}(\sigma-i\omega)|i\epsilon)}{\vartheta_{1}(\sqrt{q}(\sigma-i\omega)|i\epsilon)}
\ee
This expression vanishes as \(\sigma\goto 0\), and thus does not lead to a suitable two-point function. We thus consider the derivative
\be
\lim_{x\goto 0}\,\frac{d\,}{dx}\,
\frac{\vartheta'_{1}(x\pm iy|i\epsilon)}{\vartheta_{1}(x\pm iy|i\epsilon)}=
-i\frac{d\,}{dy}\,\frac{\vartheta'_{1}(iy|i\epsilon)}{\vartheta_{1}(iy|i\epsilon)}
\ee
Implementing the modular transformation,
\be
i\epsilon\,\frac{\vartheta'_{1}(iy|i\epsilon)}{\vartheta_{1}(iy|i\epsilon)}=
\epsilon\frac{d\,}{dy}\ln\left(\vartheta_{1}(y/\epsilon|i/\epsilon)\right)+2\pi y
\ee
we find
\be
\lim_{\sigma\goto 0}\,\frac{d\,}{d\sigma}\,\epsilon\,\hat{T}(\omega)=
-\frac{2}{\sqrt{q}}\,\epsilon\,\frac{d^2\,}{d\omega^2}
\ln\left(\vartheta_{1}(\omega\sqrt{q}/\epsilon|i/\epsilon)\right)-4\pi\sqrt{q}
\ee
For \(\epsilon\simeq 0\) with \(\omega\in\mathbb{R}\) we have
\be
\epsilon\frac{d\,}{d\omega}
\ln\left(\vartheta_{1}(\omega\sqrt{q}/\epsilon|i/\epsilon)\right)\simeq
\epsilon\frac{d\,}{d\omega}\ln\left(\sin(\pi\omega\sqrt{q}/\epsilon)\right)
=\pi\sqrt{q}\cot(\pi\omega\sqrt{q}/\epsilon)
\ee
or
\be
\lim_{\sigma\goto 0}\,\frac{d\,}{d\sigma}\,\epsilon\,\hat{T}(\omega)\simeq
-2\pi\frac{d\,}{d\omega}\cot(\pi\omega\sqrt{q}/\epsilon)-4\pi\sqrt{q}
\ee
This leads to
\be
\lim_{\sigma\goto 0}\,\frac{d\,}{d\sigma}\,\epsilon\,T
\simeq 2\pi\,\frac{\pi\sqrt{q}}{\epsilon}
\left(\csc^2(\pi\omega_{+}\sqrt{q}/\epsilon)-
\csc^2(\pi\omega_{-}\sqrt{q}/\epsilon)\right)
\ee
It does not appear that this expression can serve as a metric on fields, and thus the conclusion may be reached that the spectrum does not include operators of conformal dimension \(h<(c-1)/24\,\). Whether there is a meaningful way to introduce such operators as non-normalizable fields will not be addressed here. Since the expression for \(C_{q}\) in (\ref{k3pt}) is not analytic, it seems reasonable to conclude\footnote{See, however, the discussion in \cite{Schomerus_0306} for the \(c=1\) theory.} that only \(\alpha_j\in\mathbb{R}\) lead to well defined amplitudes.

\section{\label{strings}\large String Amplitudes}

When combined with ghost contributions, the amplitudes \(C_{\sst L}\) and \(C_{q}\) may be assembled into consistent string correlators. Choosing a unit normalization for the ghost amplitude, the corresponding three-point function is constant on the sphere and is given by
\be\label{3ptgrv}
\tilde{C}(\alpha_3,\alpha_2,\alpha_1)=
C_{q}(\alpha_3,\alpha_2,\alpha_1)\,
C_{\sst L}(\beta-\alpha_3,\beta-\alpha_2,\beta-\alpha_1|\beta)
\ee
Here \(\alpha_j\in\mathbb{R}\) and \(\beta=1/\sqrt{q}\,\). The combined central charge of the Liouville theory and its continued counterpart compensates for that of the ghost central charge\,:
\be
1+6(1/\beta+\beta)^2+1-6(1/\beta-\beta)^2=26
\ee
Similarly, the combined conformal dimension of a Liouville operator with charge \(\beta-\alpha\) and a continued operator of charge \(\alpha\) compensates for that of the ghost vertex operator\,:
\be
(\beta-\alpha)(1/\beta+\alpha)+
\alpha(1/\beta-\beta+\alpha)=1
\ee
From the expressions for \(C_{\sst L}\) (\ref{LV3pt}) and \(C_{q}\) (\ref{TL3pt}) we obtain the following simple result\,:
\be\label{3ptgrav2}
\tilde{C}(\alpha_3,\alpha_2,\alpha_1)=2\pi i\,A\,q^{-4}\,
\left(\!\rule{0mm}{4mm}\right.\!-1+[\sqrt{q}\,\hat{\alpha}]-
\sum_j\,[\sqrt{q}\,(\hat{\alpha}-2\alpha_j)]\left.\!\rule{0mm}{4mm}\right)\prod_j\;\gamma(1/q-2\alpha_{j}/\sqrt{q})
\ee
Here the constant \(A_{\sst L}\) has been absorbed into \(A\,\), and all of the \(\Upsilon\) functions have cancelled from the expression. It should be noted that if \(C_{q}\) were replaced by \(C_{\sst M}\) as in the minimal gravity of \cite{Zamolodchikov_0505}\,, only factors which depend on the normalization of vertex operators appear in the analogous string amplitude.

\section{\label{interp}\large Interpretation and Conclusion}

An interpretation of the models considered here in terms of an interacting timelike boson associated with asymptotically de Sitter cosmologies appears problematic. This is since the spectrum of primary fields does not correspond to the normalizable states of the timelike Liouville quantum mechanics treated in section \ref{tlqm} and in \cite{Kobayashi&Tsutsui_9601,Fredenhagen&Schomerus_0308}. This is further complicated by the fact that there do not appear to be normalizable states for conformal dimensions \(h<(c-1)/24\,\). In the \(c=1\) treatment of \cite{Schomerus_0306}\,, fields with these conformal dimensions are taken to be normalizable for a timelike boson since the arguments of the corresponding exponential operators are imaginary. Similarly, fields with \(h>(c-1)/24\,\) are taken to be normalizable for a spacelike boson. Through the state-operator correspondence this is also a reasonable interpretation in the case of \(c<1\,\). Having said this, it should be noted that the exponential fields only give the asymptotic form of the wavefunctions of the interacting theories, and the timelike zero-mode picture certainly leads to normalizable states with asymptotically decaying real exponentials for \(h>(c-1)/24\,\). 

However, there are reasons to suspect that the theories considered here may admit interpretations relevant to two-dimensional gravity. One of the problems with such an interpretation for generic central charge is the existence of a dual potential that appears in the Coulomb-gas computation of correlators. In Liouville theory such a potential (\(\,4\pi\tilde{\mu}\,e^{2\phi/b}\,\)) also appears as the alternative dimension one operator to the canonical Liouville potential (\(\,4\pi\mu \,e^{2b\phi}\,\)), with the respective cosmological constants being fixed with respect to one another by crossing symmetry (\ref{selfdual}). Both of these potentials admit a region of field space at weak string coupling (\(\,\phi\goto -\infty\,\)) for which a free field theory appears. For the \(c\leq 1\) timelike theory, the canonical potential (\(\,4\pi\rho\,e^{2\beta\varphi}\,\)) vanishes in the region of strong string coupling (\(\,\varphi=-i\phi\goto -\infty\,\)). For small matter conformal dimensions in a string model, a large Casimir energy prevents the field from exploring this domain. However, for sufficient matter energy it is possible for the dual potential (\(\,4\pi\tilde{\rho}\,e^{-2\varphi/\beta}\,\)) to grow large. In the two-dimensional gravity interpretation this is associated with small spatial scale in collapsing asymptotically de Sitter geometries. If the dual potential is non-zero, a free field treatment at small spatial scale is not available. It turns out that the dual cosmological constant in the timelike theory, which is also fixed with respect to the canonical cosmological constant (\ref{tlselfdual}), vanishes precisely for the central charges (\ref{topccs}) of the non-rational theories considered here (see figure 4). These are also the only central charges for which the dual cosmological constant is real. Whatever the spacetime interpretation, this is at least suggestive of a more tractable family of non-rational theories than might be expected at generic central charge.

\vspace{0.5cm}

\begin{figure}[h]
\begin{center}
\begin{picture}(200,124)
\put(6,130){\(\pi\tilde\mu\)}
\put(208,60){\(b^{-2}\)}
\includegraphics{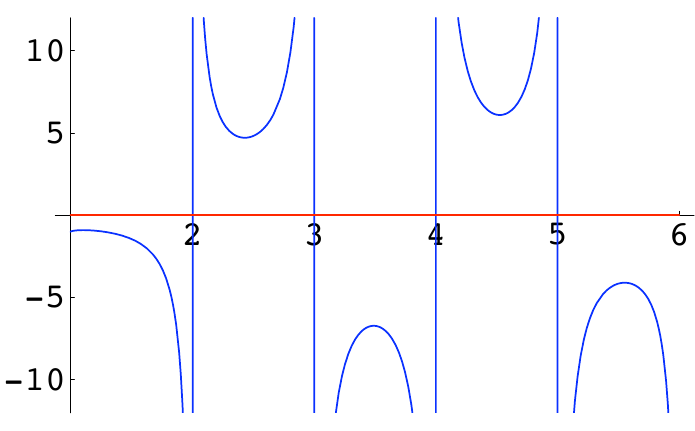}
\end{picture}
\hspace{0.5cm}
\begin{picture}(200,124)
\put(12,130){\(\pi\tilde\rho\)}
\put(208,60){\(\beta^{-2}\)}
\includegraphics{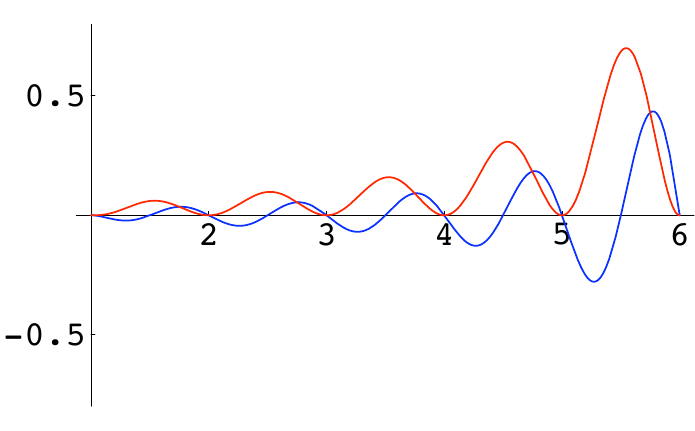}
\end{picture}
\hspace{0.5cm}
\caption{
The figure at left is the coefficient of the dual potential \(\pi\tilde{\mu}\,e^{2\phi/b}\) as a function of \(b^{-2}\) for the choice \(\pi\mu=1\) in Liouville theory. Here the real part is shown in \textcolor{blue}{blue} and the imaginary part is shown in \textcolor{red}{red}. It may be seen that the dual potential is not bounded from below for all \(c\geq 25\,\), despite the assumption of a single vertex operator per conformal dimension utilized in the conformal bootstrap. The figure at right is the corresponding coefficient of the dual potential \(\pi\tilde{\rho}\,e^{-2\varphi/\beta}\) as a function of \(\beta^{-2}\) for the choice \(\pi\rho=1\) in the continuation of Liouville theory to \(c\leq 1\). In this case the dual potential is complex and vanishes at the central charges \(c=13-6(q^{-1}+q)\) for \(q\in\mathbb{Z}_+\) of the topological minimal models.
}
\end{center}
\end{figure}

It should be said that the language of the preceding two paragraphs is largely heuristic in nature. The status of the dual potentials in the respective CFTs is well-defined from a mathematical standpoint, but their physical significance is not apparent.\footnote{This confusion exists at least in the mind of the author. For some discussion see \cite{Teschner_0104}.} Furthermore, as opposed to the continuation of the Liouville charges, the timelike rotation of the Liouville boson certainly warrants skepticism. Putting these interpretational issues aside, it does appear that the CFTs discussed here are closely associated with the non-rational \(c=1\) model of \cite{Runkel&Watts_0107,Schomerus_0306}. The extent to which these theories have physical significance is a question for further investigation.

\vspace{1cm}

{\noindent {\bf Acknowledgments}\,: The author would like to thank Emil Martinec, David Kutasov, and Ted Allen for helpful discussions.}

\end{document}